\documentclass[lettersize,onecolumn]{IEEEtran}
\usepackage{amsmath,amsfonts}
\usepackage{amssymb}
\usepackage{amsthm}
\usepackage{mathtools}
\newtheorem{theorem}{Theorem}
\newtheorem{definition}{Definition}
\newtheorem{proposition}{Proposition}
\newtheorem{corollary}{Corollary}
\newtheorem{lemma}{Lemma}
\usepackage{algorithmic}
\usepackage{algorithm}
\usepackage{array}
\usepackage[caption=false,font=normalsize,labelfont=sf,textfont=sf]{subfig}
\usepackage{textcomp}
\usepackage{stfloats}
\usepackage{booktabs}
\usepackage{url}
\usepackage{verbatim}
\usepackage{graphicx}
\usepackage{cite}
\usepackage{centernot}
\begin{document}
\bstctlcite{IEEEtran:BSTcontrol}

\title{Queue-Dynamic Certification and\\
Monotone Refinement of Stochastic Bounds}

\author{Weixuan~Fan,
        Jin~Li,
        Yonghan~Wu,
        Yi~Huang, and~Min~Zhang%
\thanks{Weixuan Fan, Yi Huang, and Min Zhang are with the Department of
Electronic Engineering, Beijing University of Posts and Telecommunications,
Beijing 100080, China (e-mail: {\upshape weixuan\_fan@bupt.edu.cn;
hyi@bupt.edu.cn; mzhang@bupt.edu.cn}).}%
\thanks{Jin Li is with the School of Optoelectronic Science and Engineering,
South China Normal University, Guangzhou 510006, China
(e-mail: {\upshape lijin930629@163.com}).}%
\thanks{Yonghan Wu is with the Department of Broadband Communication,
Peng Cheng Laboratory, Shenzhen 518000, China
(e-mail: {\upshape wuyh02@pcl.ac.cn}).}}

\maketitle

\begin{abstract}
Predictive performance bounds support resource provisioning for statistical quality of service. Under imperfect traffic and service models and finite
packet streams, adjustable conservatism helps balance safety margins and resource efficiency. However, optimizing a bound within a prescribed family generally offers no systematic way to tighten it further toward the stationary tail. For stable stationary queues with a sufficient predictive state, we derive a positive affine workload-tail operator from the exact reflected block
recursion. We complete a valid state-conditioned workload bound to its least dominating supersolution, from which operator iteration generates a monotone hierarchy of valid bounds without requiring the stationary tail in advance. We
further establish a verifiable input condition under which completion preserves tail decay and refinement converges to the stationary workload tail, even without uniform decay across states. Interpolating adjacent levels makes
conservatism continuously selectable: every selected bound remains a supersolution and can be tightened further by the same operator. Stationary averaging yields unconditional workload bounds. Exact packet-Palm transfer
preserves the hierarchy for packet-completion delay violation probabilities and relates the finite-depth delay gap directly to the remaining workload error. We evaluate the framework across diverse wireless queues and illustrate
how refinement depth governs model-based safety margins.
\end{abstract}


\section{Introduction}
\label{sec:introduction}

Statistical quality-of-service (QoS) bounds support resource provisioning by relating statistical performance metrics, such as the workload tail and the delay violation probability (DVP), to reliability requirements.
Extensive foundational research has established analytical approaches for deriving such stochastic performance bounds.
Effective bandwidth~\cite{chang1995effectivebandwidth} and effective capacity~\cite{wu2003effectivecapacity} connect traffic and wireless-service statistics to queueing performance through logarithmic exponential transforms.
Moment-generating-function (MGF)-based queueing analysis~\cite{chang1994stability} and stochastic network calculus (SNC)~\cite{jiang2006basic} derive probabilistic workload and delay guarantees. Exponential martingale methods, rooted in classical queueing
inequalities~\cite{kingman1964martingale} and developed for communication scheduling~\cite{POLOCZEK201456}, can substantially tighten conventional Chernoff and union-bound results for many light-tailed independent or short-range-dependent models. Service-martingales~\cite{7218466} extend this methodology to random service and yield sharp delay bounds for Markovian sources over random-access channels. Stochastic burstiness bounds~\cite{starobinski2000stochastically} extend traffic
descriptions beyond exponential envelopes. Statistical envelopes and sample-path bounds further enable non-asymptotic delay analysis for heavy-tailed and
self-similar traffic~\cite{liebeherr2012delay}. These methods inform practical decisions, including QoS-constrained radio resource allocation~\cite{zhu2024qos}
and computational offloading~\cite{picano2024combined}.

Theoretical research treats improved tightness as a principal measure of progress: tighter bounds reduce overprovisioning under a correctly specified model.
However, deployment also faces imperfect traffic and
service descriptions, while empirical violation fractions remain random over finite packet streams. Measurement-based network calculus incorporates arrival-estimation uncertainty~\cite{beck2014statistical}; our recent digital-twin framework combines parallel simulations, Bayesian inference, and parameter extrapolation to estimate finite-window DVP distributions and model-based upper quantiles~\cite{fan2026digitaltwin}. These concerns motivate selectable analytical conservatism, allowing resource provisioning to retain safety margins appropriate to operating conditions while curbing unnecessary
overprovisioning.

Established analytical methods typically tighten a bound by optimizing parameters within a prescribed family. Even the best bound in that family can remain above the exact stationary tail. Recent work has approached bound improvement in several ways. Ciucu et al.~\cite{ciucu2024ultrasharp} use exponential tilting and an overshoot series to obtain progressively sharper bounds for GI/G/1, alternating-renewal, and Markov-fluid queues. Retaining more terms trades computation for tightness, with the refinement tied to the overshoot representation available for each model. In a related fixed-point approach for tandem queues, Ciucu and Mehri~\cite{ciucu2025tandem} construct comparison functions that satisfy an integral inequality for a joint
distribution, yielding polynomial-exponential bounds on waiting and sojourn times. In deterministic network calculus, Tabatabaee and Le
Boudec~\cite{tabatabaee2022deficit} iteratively strengthen strict service curves for Deficit Round-Robin to improve worst-case delay guarantees. These advances leave open how to turn an independently derived valid workload bound into a starting point from which the exact queue dynamics generate a monotone hierarchy of valid bounds approaching the stationary tail. Translating these workload bounds into packet-level guarantees also requires an explicit change of sampling law, from observation times for workload and virtual delay to arriving packets for completion delay.

To address these issues, we develop a framework grounded in exact queue dynamics for stable stationary queues. A sufficient predictive state determines the joint conditional law of the block net input, the restart workload (the workload at the end of a block started empty), and the next predictive state. The net input and restart workload yield an exact reflected block recursion and an affine tail operator $\mathcal T f=a+\mathcal Kf$ that separates violations generated within a block from those that depend on earlier workload. The stationary conditional workload tail $Q$ is
a fixed point of this operator.

Monotone tightening of a state-conditioned bound $F\geq Q$ under $\mathcal T$ requires $\mathcal TF\leq F$. We therefore complete $F$ to its least dominating supersolution $U_0$, with an equivalent linear program (LP) when an exact finite-dimensional realization is available. Starting from $U_0$, the iteration $U_{n+1}=\mathcal TU_n$ ensures $Q\leq U_{n+1}\leq U_n$ at every depth.
Interpolating adjacent levels makes conservatism continuously selectable while keeping each selected bound a supersolution. The original operator can therefore continue tightening it without taking a pointwise minimum at each step. For stationary ergodic net input with a finite first absolute moment and negative drift, completion preserves statewise tail decay and refinement converges to $Q$ under an integrability condition on $F$.
For each level in a sequence decreasing to zero, the state-dependent workload threshold at which $F$ reaches or falls below that level must have finite mean under the stationary state law. This condition holds in our queue with packetized traffic driven by a Gaussian autoregressive (AR) source of order $q$, where the conditional workload tail has essential
supremum one over predictive states at every finite workload threshold.

We then obtain unconditional workload bounds by stationary averaging and packet-completion DVP bounds by exact packet-Palm transfer. Both operations preserve validity, monotonicity, and convergence, and the Palm transfer maps the propagated workload error exactly to the packet DVP gap.

We derive state-conditioned input bounds under Markov-modulated wireless service for Markov-modulated compound-Poisson (MMCPP) arrivals, AR-driven packetized traffic, and Poisson--Pareto batch arrivals. Together with the
independent-block GI/GI/1-type baseline, these models illustrate completion, workload-to-packet transfer, and how resource allocation, arrival dependence, and heavy-tailed bursts affect refinement depth.

The rest of this paper is organized as follows.
Section~\ref{sec:system_model} introduces the model and problem. Sections~\ref{sec:workload_refinement} and~\ref{sec:palm_transfer} develop
workload refinement and packet-Palm transfer, respectively. Section~\ref{sec:numerical_validation} presents validation, with proofs of
the model-specific state-conditioned bounds in the appendices.
Section~\ref{sec:conclusion} concludes the paper.

\section{System Model and Problem Formulation}
\label{sec:system_model}

\subsection{Block-Based Single-Hop Service Queue}
\label{subsec:queue_system}

We consider a stable stationary single-hop service queue operating under a
work-conserving first-in--first-out (FIFO) discipline. Arrivals and service
are measured in a common workload unit, such as bits for transmission or
computational work for task processing. This process-level formulation
accommodates wired transmission, wireless transmission, and computation
queues with the same workload-balance dynamics.
Time is partitioned into observation blocks $(kH,(k+1)H]$, indexed by
$k\in\mathbb{Z}$. The observation grid may align with the physical service
slots or use a different partition. Within block $k$, let
$A_k(u)=A(kH,kH+u]$ and $S_k(u)=S(kH,kH+u]$, $0\leq u\leq H$, denote
the cumulative arrived workload and available service, with
$A_k(0)=S_k(0)=0$.

Define the within-block net-input process $X_k(u)=A_k(u)-S_k(u)$, its endpoint
increment $Y_k$, and the restart term $R_k$ by
\begin{equation}
\begin{aligned}
    X_k(u) &\triangleq A_k(u)-S_k(u), \qquad 0\leq u\leq H,\\
    Y_k &\triangleq X_k(H),\\
    R_k &\triangleq X_k(H)-\inf_{0\leq u\leq H}X_k(u).
\end{aligned}
\label{eq:block_quantities}
\end{equation}
Here $Y_k$ is the net input over the complete block and $R_k$ is the
workload at its end when the block starts empty.

Let $W_k=W(kH)$ denote the right-continuous workload at the start of
block $k$, including any batch arriving at $kH$.
The pathwise within-block workload balance gives the exact block-boundary
recursion
\begin{equation}
    W_{k+1}
    =
    \max\left\{
        W_k+Y_k,\,
        R_k
    \right\}.
    \label{eq:reflected_block_recursion}
\end{equation}
Since $R_k\geq0$ and $R_k-Y_k\geq0$, this is equivalently
$W_{k+1}=R_k+(W_k-(R_k-Y_k))^+$, a Lindley-type
recursion~\cite{boxma2023fluctuation}.
The restart term is needed because the queue may empty within the block.

We use the finite causal stationary workload obtained by starting the queue
empty in the remote past,
\[
 W_k=\sup_{m\geq1}\left\{R_{k-m}+\sum_{\ell=1}^{m-1}Y_{k-\ell}\right\}.
\]

For any threshold $x\geq0$, \eqref{eq:reflected_block_recursion} immediately
gives the disjoint decomposition
\begin{equation}
    \{W_{k+1}>x\}
    =
    \{R_k>x\}
    \cup
    \{R_k\leq x,\;W_k>x-Y_k\}.
    \label{eq:block_violation_decomposition}
\end{equation}
The first event is generated within the block; the second depends on the
incoming workload.

To accommodate temporal dependence, let $Z_k\in\mathsf Z$ be the predictive
state, with $\mathsf Z$ a standard Borel space. The block-mark history
available at the start of block $k$ is
$\mathcal F_k^-:=\sigma\bigl((Y_j,R_j,Z_j)_{j<k},Z_k\bigr)$.
We assume that the marked sequence $(Z_k,Y_k,R_k)_{k\in\mathbb Z}$ is
jointly stationary and that $Z_k$ is sufficient for the next marked
transition: for each measurable rectangle
$B=B_Y\times B_R\times B_Z$ in
$\mathbb R\times[0,\infty)\times\mathsf Z$,
\begin{equation}
 \mathbb P\bigl((Y_k,R_k,Z_{k+1})\in B\mid\mathcal F_k^-\bigr)
 =\mathsf K(B\mid Z_k)
 \qquad\text{a.s.}
    \label{eq:marked_predictive_kernel}
\end{equation}
Here $\mathsf K$ is the joint conditional kernel of the block mark and the
next state given $Z_k$. The assumption permits dependence among
$Y_k$, $R_k$, and $Z_{k+1}$ and covers discrete, continuous, and mixed
predictive states. The causal representation makes $W_k$
$\mathcal F_k^-$-measurable.

\subsection{Problem Formulation}
\label{subsec:packet_delay_problem}

We consider a stochastic arrival process in which packets arrive individually
or in batches. Let $t_b$ denote the arrival epoch of batch $b$ and $N_b\geq1$ its packet count. A single-packet arrival is the special case $N_b=1$.
Let $L_{b,r}>0$ be the length of packet $r$ in batch $b$ and
$L_{b,\leq r}:=\sum_{\ell=1}^{r}L_{b,\ell}$ its inclusive prefix workload.
The total batch workload $L_{b,\leq N_b}$ enters as one jump. With
$W(t_b^-)$ the workload immediately before the batch, the work required
to complete packet $r$ is
\begin{equation}
    w_{b,r}^{\mathrm{req}}
    \triangleq
    W(t_b^-)+L_{b,\leq r}.
    \label{eq:tagged_packet_required_work}
\end{equation}
Index packets globally by $i$ in arrival and batch FIFO order, with
$T_i=t_b$, $L_i=L_{b,r}$, and $w_i^{\mathrm{req}}=w_{b,r}^{\mathrm{req}}$
for packet $(b,r)$.
Let $S(s,t)$ denote the cumulative service available over $(s,t]$.
With continuous-time fluid service, packet $i$ completes when the service
available since its arrival first reaches $w_i^{\mathrm{req}}$. Its
completion time and delay are
\begin{equation}
\begin{aligned}
    C_i
    &\triangleq
    \inf\{t\geq T_i:S(T_i,t)\geq w_i^{\mathrm{req}}\},\\
    D_i
    &\triangleq
    C_i-T_i .
\end{aligned}
\label{eq:packet_completion_delay}
\end{equation}

At an arbitrary time $t$, the virtual delay is the time required to clear
the workload present in the queue,
\begin{equation}
    V(t)
    \triangleq
    \inf\{u\geq0:S(t,t+u)\geq W(t)\}.
    \label{eq:virtual_delay}
\end{equation}
The virtual-delay tail samples the queue in time. By comparison, the packet-completion tail samples arrivals and uses the FIFO workload through the tagged packet.

For a delay threshold $d>0$, we define the stationary packet completion-delay
violation probability as the long-run fraction of violating packets,
\begin{equation}
    p_{\rm pkt}(d)
    \triangleq
    \lim_{N\rightarrow\infty}
    \frac{1}{N}
    \sum_{i=1}^{N}
    \mathbf 1_{\{D_i>d\}},
    \label{eq:packet_dvp}
\end{equation}
under the assumed stationary ergodic regime. This packet-level quantity is
the target performance metric of this paper.

Exact evaluation of $p_{\rm pkt}(d)$ can be difficult when the arrival
and service processes induce a high-dimensional, continuous, or otherwise
complex predictive state. In communication-system design, however, resource
provisioning is often based on a reliability requirement
$p_{\rm pkt}(d)\leq\varepsilon$. A tractable probabilistic upper bound is
therefore useful because it provides an explicit analytical safety margin,
whereas operating directly at a model-predicted exact-performance boundary
leaves no such margin.

Existing analytical workload bounds provide starting points whose conservatism
we seek to reduce systematically. Let $F$ denote a valid state-conditioned
analytical workload bound. Our objective is to construct its supersolution
completion $U_0\geq F$ and, from this certified starting point, a hierarchy
$\{U_n\}_{n\geq0}$ satisfying
\begin{equation}
    Q\leq U_{n+1}\leq U_n,
    \qquad
    U_n\downarrow Q,
\label{eq:state_hierarchy_goal}
\end{equation}
where $Q$ denotes the stationary conditional workload tail. The goal is to
select a degree of conservatism and then continue tightening the chosen bound
through the same queue dynamics.

The final objective is to transfer this monotone state-conditioned hierarchy
to packet completion delay and obtain bounds $B_n(d)$ satisfying
\begin{equation}
    p_{\rm pkt}(d)
    \leq
    B_{n+1}(d)
    \leq
    B_n(d),
    \qquad
    B_n(d)\downarrow p_{\rm pkt}(d).
    \label{eq:packet_hierarchy_goal}
\end{equation}

\section{Stationary Workload Characterization and Queue-Dynamic Bound Refinement}
\label{sec:workload_refinement}

\subsection{Exact Reflected Workload Operator}
\label{subsec:exact_workload_operator}

Let
\begin{equation}
    Q(x,z)
    \triangleq
    \mathbb P(W_k>x\mid Z_k=z),
    \qquad x\geq0,
    \label{eq:conditional_workload_tail}
\end{equation}
denote the stationary workload tail conditioned on the predictive state.

The forward marked kernel introduced in
Section~\ref{sec:system_model} specifies the one-block evolution of the
predictive state. To characterize the workload conditioned on the current
state, define its stationary reverse counterpart by
\begin{equation}
\begin{aligned}
    \check{\mathsf K}(B_Y,B_R,B_Z\mid z')
    \triangleq
    \mathbb P\bigl(
        &Y_{k-1}\in B_Y,\,
        R_{k-1}\in B_R,\,
        Z_{k-1}\in B_Z
        \mid Z_k=z'
    \bigr),
\end{aligned}
\label{eq:reverse_marked_kernel}
\end{equation}
for measurable sets
$B_Y\in\mathcal B(\mathbb R)$,
$B_R\in\mathcal B(\mathbb R_+)$, and
$B_Z\in\mathcal B(\mathsf Z)$.
Let $\pi$ be the stationary law of $Z_k$. The stationary joint law
determines the reverse kernel up to $\pi$-null sets.
Iterating \eqref{eq:marked_predictive_kernel} shows that, given a boundary
state, the conditional law of any finite segment of subsequent marked
transitions does not depend on the preceding block-mark history. Hence the
next earlier mark and state have law $\check{\mathsf K}$ conditional on the
current boundary state, even given the more recent blocks.

For the regular conditional laws, choose a common full-$\pi$-measure set
of states that works simultaneously for all workload thresholds.
By stationarity, retaining the states from which the reverse chain stays
in that set almost surely at all depths yields a reverse-closed
full-$\pi$-measure set, denoted by $\mathsf Z_*$. Throughout this section,
pointwise comparisons and state-conditioned conclusions are understood
on $[0,\infty)\times\mathsf Z_*$, with $\mathsf Z_*$ refined to a smaller
reverse-closed full-measure set when additional conditional laws are used.

For a measurable function
$f:[0,\infty)\times\mathsf Z\rightarrow[0,1]$, define
\begin{equation}
\begin{aligned}
    (\mathcal T f)(x,z')
    \triangleq
    \mathbb E\Big[
        &\mathbf 1_{\{R_{k-1}>x\}}
        +
        \mathbf 1_{\{R_{k-1}\leq x\}}
        f(x-Y_{k-1},Z_{k-1})
        \,\Big|\, Z_k=z'
    \Big],
\end{aligned}
\label{eq:exact_queue_operator}
\end{equation}
for $x\geq0$, using the reverse marked kernel
\eqref{eq:reverse_marked_kernel}. Since $Y_{k-1}\leq R_{k-1}$, the continuation event
$R_{k-1}\leq x$ ensures $x-Y_{k-1}\geq0$; the indicator-weighted tail
evaluation is taken only on this event.

This operator follows directly from the reflected block recursion
\[
    W_k
    =
    \max\{W_{k-1}+Y_{k-1},R_{k-1}\}.
\]
If $R_{k-1}>x$, the threshold is violated independently of the incoming
workload. Otherwise, violation occurs precisely when
\[
    W_{k-1}
    >
    x-Y_{k-1}.
\]

Define
\begin{equation}
\begin{aligned}
    a(x,z')
    &\triangleq
    \mathbb P(R_{k-1}>x\mid Z_k=z'),
    \\
    (\mathcal Kf)(x,z')
    &\triangleq
    \mathbb E\!\left[
        \mathbf 1_{\{R_{k-1}\leq x\}}
        f(x-Y_{k-1},Z_{k-1})
        \mid Z_k=z'
    \right].
\end{aligned}
\label{eq:affine_operator_components}
\end{equation}
Then
\begin{equation}
    \mathcal Tf
    =
    a+\mathcal Kf.
\label{eq:affine_queue_operator}
\end{equation}
The term $a$ accounts for violations generated within the most recent block,
whereas $\mathcal K$ propagates dependence on earlier workload.
The query $f(x-Y_{k-1},Z_{k-1})$ pairs the threshold shifted by the
preceding block's net input with the workload tail conditioned on that
block's initial state. The reverse kernel carries this dependence into
each update.
For independent and identically distributed (i.i.d.) block marks and a
singleton predictive state, this reduces to the perturbed-random-walk tail
equation in~\cite{araman2006tail}, whose source term $b$ and linear
operator $T$ correspond to $a$ and $\mathcal K$.
The same expression canonically extends $\mathcal K$ to bounded signed
functions, which will be used below for difference and residual identities.

\begin{lemma}[Order Properties of the Reflected Workload Operator]
\label{lem:T_order_properties}
The operator $\mathcal T$ is monotone:
\begin{equation}
    f\leq g
    \quad\Longrightarrow\quad
    \mathcal Tf\leq\mathcal Tg.
    \label{eq:T_monotonicity}
\end{equation}
If $0\leq f\leq1$, then
\[
    0\leq\mathcal Tf\leq1.
\]
In particular, the constant function $\mathbf1(x,z)\equiv1$ satisfies
\begin{equation}
    \mathcal T\mathbf1=\mathbf1.
    \label{eq:T_one_fixed}
\end{equation}
Moreover, if $f_r\uparrow f$ pointwise, then
\begin{equation}
    \mathcal Tf_r
    \uparrow
    \mathcal Tf.
    \label{eq:T_continuity_below}
\end{equation}
\end{lemma}

\begin{proof}
Since $\mathcal K$ is a positive linear operator,
$f\leq g$ implies
$\mathcal Kf\leq\mathcal Kg$, and therefore
$\mathcal Tf\leq\mathcal Tg$.

If $0\leq f\leq1$, the random variable inside
\eqref{eq:exact_queue_operator} lies in $[0,1]$, which proves
$0\leq\mathcal Tf\leq1$.
For $f=\mathbf1$, the two indicator terms sum to one, proving
\eqref{eq:T_one_fixed}.

Finally, if $f_r\uparrow f$, the corresponding integrands in
\eqref{eq:exact_queue_operator} increase pointwise.
The monotone convergence theorem yields
\eqref{eq:T_continuity_below}.
\end{proof}

\begin{theorem}[Exact Stationary Workload Fixed Point]
\label{thm:exact_workload_fixed_point}
For a stable stationary queue with the predictive-state representation of
Section~\ref{sec:system_model}, the conditional stationary workload tail
satisfies
\begin{equation}
    Q
    =
    \mathcal TQ
    =
    a+\mathcal KQ.
\label{eq:exact_workload_fixed_point}
\end{equation}
\end{theorem}

\begin{proof}
For $x\geq0$, the event $\{W_k>x\}$ is the disjoint union of
$\{R_{k-1}>x\}$ and
$\{R_{k-1}\leq x,\,W_{k-1}>x-Y_{k-1}\}$.
By $\mathcal F_{k-1}^-$-measurability of $W_{k-1}$ and
\eqref{eq:marked_predictive_kernel}, its conditional law given $Z_{k-1}$
is unchanged by additionally conditioning on $(Y_{k-1},R_{k-1},Z_k)$.
Evaluating this law at $x-Y_{k-1}$ on the continuation event and
conditioning on $Z_k=z'$ gives
\[
\begin{aligned}
    Q(x,z')
    =
    \mathbb E\Big[
        &\mathbf1_{\{R_{k-1}>x\}}
        +
        \mathbf1_{\{R_{k-1}\leq x\}}
        Q(x-Y_{k-1},Z_{k-1})
        \,\Big|\,Z_k=z'
    \Big],
\end{aligned}
\]
which is exactly $\mathcal TQ$.
\end{proof}

\subsection{Analytical Bounds and Supersolution Completion}
\label{subsec:supersolution}

Suppose an analytical argument provides a measurable workload-tail envelope
$F:[0,\infty)\times\mathsf Z\rightarrow[0,1]$ satisfying
\begin{equation}
    Q(x,z)\leq F(x,z)\leq1,
    \qquad x\geq0,
    \label{eq:analytical_bound_def}
\end{equation}
and, for every $z\in\mathsf Z$,
\begin{equation}
    F(x_1,z)\geq F(x_2,z),
    \qquad
    0\leq x_1\leq x_2,
    \label{eq:analytical_bound_monotone}
\end{equation}
with
\begin{equation}
    \lim_{x\rightarrow\infty}F(x,z)=0.
    \label{eq:analytical_bound_tail}
\end{equation}
We call such an $F$ an \emph{analytical workload bound}.
While stationary-average workload bounds~\cite{POLOCZEK201456,10.1145/3287322}
do not generally ensure \eqref{eq:analytical_bound_def} on $\mathsf Z_*$,
a common envelope valid on $\mathsf Z_*$ is admissible.

\begin{definition}[Supersolution]
A measurable function
$U:[0,\infty)\times\mathsf Z\rightarrow[0,1]$
is a \emph{supersolution} of the stationary workload equation if
\begin{equation}
    \mathcal TU\leq U.
    \label{eq:supersolution_def}
\end{equation}
\end{definition}

The condition $Q\leq F$ does not by itself imply
$\mathcal TF\leq F$.  To see the difference, define the slack
$E\triangleq F-Q$.

Since $Q=\mathcal{T}Q$, we have
\begin{align}
    \mathcal{T}F-F
    &=
    \bigl(\mathcal{T}F-\mathcal{T}Q\bigr)
    -
    (F-Q)
    \nonumber\\
    &=
    \mathcal{K}(F-Q)-(F-Q)
    \nonumber\\
    &=
    \mathcal{K}E-E,
    \label{eq:supersolution_slack_identity}
\end{align}
where the second equality follows from the affine representation
$\mathcal{T}f=a+\mathcal{K}f$ in
\eqref{eq:affine_queue_operator} and the linearity of $\mathcal{K}$.
Hence, for every state--threshold pair $(x,z)$,
\begin{equation}
    (\mathcal TF)(x,z)\leq F(x,z)
    \quad\Longleftrightarrow\quad
    (\mathcal KE)(x,z)\leq E(x,z).
    \label{eq:pointwise_slack_equivalence}
\end{equation}
Therefore
\begin{equation}
    \mathcal TF\leq F
    \quad\Longleftrightarrow\quad
    \mathcal KE\leq E.
\label{eq:global_slack_equivalence}
\end{equation}

Thus $E\geq0$ expresses only that $F$ is an upper bound, whereas the
supersolution property additionally requires the slack to be superharmonic
with respect to the continuation operator $\mathcal K$.  In the present
operator sense, this means that one-step propagated slack cannot exceed the
slack already available at the current state--threshold pair.

If this condition fails, then there exists some $(x^\star,z^\star)$ such that
$(\mathcal KE)(x^\star,z^\star)>E(x^\star,z^\star)$.  By
\eqref{eq:supersolution_slack_identity},
\begin{equation}
    (\mathcal TF)(x^\star,z^\star)>F(x^\star,z^\star),
    \label{eq:analytical_bound_positive_defect}
\end{equation}
so $F$ is not a supersolution despite remaining a valid upper bound on $Q$.

To repair the positive violations of the supersolution inequality, define
the completion map $\Phi(f):=f\vee\mathcal Tf$, where $\vee$ denotes
pointwise maximum. Starting from $C_0=F$, iterate
\begin{equation}
    C_{r+1}=\Phi(C_r)=C_r\vee\mathcal TC_r,
    \qquad r\geq0,
\label{eq:workload_completion_boxed}
\end{equation}
where $r$ denotes the completion index. Each step adds exactly
the positive supersolution defect:
\begin{equation}
    C_{r+1}-C_r=[\mathcal TC_r-C_r]^+,
\label{eq:completion_defect_correction}
\end{equation}
with $[g]^+=\max\{g,0\}$ pointwise. Repeated updates allow a lift to
propagate through the state--threshold dependencies in $\mathcal T$.
The following proposition identifies the limit as the least
supersolution dominating $F$.

\begin{proposition}[Least-Supersolution Completion]
\label{prop:supersolution_completion}
Let $F$ satisfy \eqref{eq:analytical_bound_def}, and define
$C_0=F$ and $C_{r+1}=\Phi(C_r)$.  Then
\begin{equation}
    Q\leq C_0\leq C_1\leq C_2\leq\cdots,
    \label{eq:completion_monotone}
\end{equation}
with $0\leq C_r\leq1$ for every $r$.  Hence
\begin{equation}
    U_0
    \triangleq
    \lim_{r\to\infty}C_r
    \label{eq:completion_limit}
\end{equation}
exists pointwise and satisfies
\begin{equation}
    \mathcal TU_0\leq U_0,
    \qquad
    U_0\geq F.
\label{eq:completion_supersolution}
\end{equation}
Moreover, $U_0$ is the least supersolution lying above $F$:
\begin{equation}
    U_0
    =
    \inf\left\{
        U:
        U\geq F,\;
        \mathcal TU\leq U
    \right\}.
\label{eq:least_supersolution}
\end{equation}
\end{proposition}

\begin{proof}
The monotonicity of the completion sequence is immediate from
$C_{r+1}=C_r\vee\mathcal TC_r$, since $C_{r+1}\geq C_r$.

The lower bound $Q\leq C_r$ follows by induction.  It holds at $r=0$.  If
$Q\leq C_r$, Lemma~\ref{lem:T_order_properties} gives
$Q=\mathcal TQ\leq\mathcal TC_r\leq C_{r+1}$.

Similarly, if $0\leq C_r\leq1$, then
$0\leq\mathcal TC_r\leq1$, so
$0\leq C_{r+1}\leq1$.  Hence $\{C_r\}$ is pointwise nondecreasing and
bounded, and the pointwise limit $U_0$ exists.

Since $C_r\uparrow U_0$, continuity from below of $\mathcal T$ gives
$\mathcal TC_r\uparrow\mathcal TU_0$.  Therefore
\begin{align}
    U_0
    &=
    \lim_{r\to\infty}C_{r+1}
    \nonumber\\
    &=
    \lim_{r\to\infty}
    \left(C_r\vee\mathcal TC_r\right)
    \nonumber\\
    &=
    \left(\lim_{r\to\infty}C_r\right)
    \vee
    \left(\lim_{r\to\infty}\mathcal TC_r\right)
    \nonumber\\
    &=
    U_0\vee\mathcal TU_0.
    \label{eq:completion_limit_chain}
\end{align}
Thus $\mathcal TU_0\leq U_0$.  Since $C_r\geq C_0=F$, also $U_0\geq F$.

It remains to prove minimality.  Let $U$ be any supersolution satisfying
$U\geq F$.  We show by induction that $C_r\leq U$ for all $r$.  The claim
holds at $r=0$ because $C_0=F\leq U$.  If $C_r\leq U$, monotonicity of
$\mathcal T$ gives
$\mathcal TC_r\leq\mathcal TU\leq U$.  Hence
\[
    C_{r+1}
    =
    C_r\vee\mathcal TC_r
    \leq U.
\]
Taking the pointwise limit yields $U_0\leq U$.  Since $U_0$ itself is a
supersolution above $F$, it is the least such supersolution.
\end{proof}

We denote this completion by
\begin{equation}
    \mathcal C(F)
    \triangleq
    U_0.
    \label{eq:completion_operator}
\end{equation}

\begin{proposition}[Direct Characterization of the Completion Limit]
\label{prop:direct_completion_limit}
Under the assumptions of
Proposition~\ref{prop:supersolution_completion}, let
$C_0=F$ and
\[
    C_{r+1}
    =
    C_r\vee\mathcal{T}C_r,
    \qquad r\geq0,
\]
and let $U_0=\lim_{r\to\infty}C_r$ denote the least-supersolution completion
of $F$.  Then the completion limit admits the direct fixed-point
characterization
\begin{equation}
    U_0
    =
    F\vee\mathcal{T}U_0.
\label{eq:completion_fixed_point}
\end{equation}
Equivalently, $U_0$ satisfies the pointwise complementarity conditions
\begin{equation}
\begin{aligned}
    U_0-F &\geq 0,\\
    U_0-\mathcal{T}U_0 &\geq 0,\\
    \min\!\left\{
        U_0-F,\,
        U_0-\mathcal{T}U_0
    \right\} &=0.
\end{aligned}
\label{eq:completion_complementarity}
\end{equation}
Moreover,
\begin{equation}
    U_0
    =
    \inf
    \left\{
        U:
        U\geq F,\;
        \mathcal{T}U\leq U
    \right\},
\label{eq:direct_least_supersolution}
\end{equation}
where the infimum is understood pointwise.  Hence $U_0$ can be
characterized without reference to any finite completion depth.
\end{proposition}

\begin{proof}
We first show that, along the completion sequence initialized by $C_0=F$,
the recursion can equivalently be written as
\begin{equation}
    C_{r+1}
    =
    F\vee\mathcal{T}C_r.
    \label{eq:completion_F_fixed_point_iteration}
\end{equation}
For $r=0$, this follows immediately from
$C_1=F\vee\mathcal{T}F$.

Suppose that
$C_r=F\vee\mathcal{T}C_{r-1}$.
Since the completion sequence is nondecreasing,
$C_r\geq C_{r-1}$, and monotonicity of $\mathcal{T}$ gives
$\mathcal{T}C_r\geq\mathcal{T}C_{r-1}$.  Therefore,
\[
    C_r
    =
    F\vee\mathcal{T}C_{r-1}
    \leq
    F\vee\mathcal{T}C_r.
\]
It follows that
\[
    C_r\vee\mathcal{T}C_r
    \leq
    F\vee\mathcal{T}C_r.
\]
Conversely, since $C_r\geq F$,
\[
    F\vee\mathcal{T}C_r
    \leq
    C_r\vee\mathcal{T}C_r.
\]
Hence \eqref{eq:completion_F_fixed_point_iteration} holds for every $r$.
This is the finite-horizon Snell-envelope recursion~\cite{peskir2006optimal}
for the backward state--threshold process, with stopping payoff $F$
before absorption and unit payoff upon a within-block violation.

By Proposition~\ref{prop:supersolution_completion},
$C_r\uparrow U_0$.  Continuity from below of $\mathcal{T}$, as stated in
\eqref{eq:T_continuity_below} of Lemma~\ref{lem:T_order_properties}, yields
$\mathcal{T}C_r\uparrow\mathcal{T}U_0$.  Taking the pointwise limit in
\eqref{eq:completion_F_fixed_point_iteration} therefore gives
\[
    U_0
    =
    F\vee\mathcal{T}U_0,
\]
which proves \eqref{eq:completion_fixed_point}.

The fixed-point relation
\eqref{eq:completion_fixed_point}
is equivalent to
$U_0\geq F$ and $U_0\geq\mathcal{T}U_0$, together with the condition that
at least one of the two inequalities is active at each state--threshold
pair.  This gives \eqref{eq:completion_complementarity}.

Finally, Proposition~\ref{prop:supersolution_completion} already establishes
that $U_0$ is dominated by every supersolution $U$ satisfying $U\geq F$.
Hence \eqref{eq:direct_least_supersolution} follows.
\end{proof}

For finite-dimensional affine operators, we can compute the least
dominating supersolution directly by solving an LP. We collect the
represented state--threshold values into a vector and solve for them jointly.

Let $D$ denote the number of represented state--threshold coordinates,
$\{(x_\nu,z_\nu)\}_{\nu=1}^{D}$, and write
\begin{equation}
    \mathbf u
    :=\bigl(U(x_\nu,z_\nu)\bigr)_{\nu=1}^{D}
    \in[0,1]^D,
\label{eq:finite_tail_vector}
\end{equation}
with analytical-bound vector
\begin{equation}
    \mathbf f
    :=\bigl(F(x_\nu,z_\nu)\bigr)_{\nu=1}^{D}.
\label{eq:finite_bound_vector}
\end{equation}
Consider a finite-dimensional affine operator
\begin{equation}
    \mathbf T(\mathbf u)=\mathbf a+\mathbf K\mathbf u,
\label{eq:finite_affine_operator}
\end{equation}
where $\mathbf a\in\mathbb R_+^D$, $\mathbf K\in\mathbb R_+^{D\times D}$,
and $\mathbf T$ maps $[0,1]^D$ into itself.

\begin{corollary}[Direct Finite-Dimensional Completion]
\label{cor:direct_finite_completion}

Let $\mathbf I$ denote the $D\times D$ identity matrix and
$\mathbf1\in\mathbb R^D$ the all-one vector.  Define
\begin{equation}
    \mathcal S_{\mathbf f}
    \triangleq
    \left\{
        \mathbf u\in[0,1]^D:
        \mathbf u\geq\mathbf f,\;
        (\mathbf I-\mathbf K)\mathbf u\geq\mathbf a
    \right\}.
\label{eq:finite_supersolution_set}
\end{equation}

The finite-dimensional completion sequence
\begin{equation}
\begin{aligned}
    \mathbf c_0&=\mathbf f,\qquad
    \mathbf c_{r+1}=\mathbf c_r\vee\mathbf T(\mathbf c_r),\\
    \mathbf u_{0,F}&:=\lim_{r\to\infty}\mathbf c_r
\end{aligned}
\label{eq:finite_completion_limit}
\end{equation}
has a limit equal to the componentwise least element of
$\mathcal S_{\mathbf f}$:
\begin{equation}
    \mathbf u_{0,F}
    \leq
    \mathbf u,
    \qquad
    \forall\,\mathbf u\in
    \mathcal S_{\mathbf f},
\label{eq:finite_componentwise_least}
\end{equation}
where vector inequalities are understood componentwise.

Moreover, for any $\mathbf w\in\mathbb R_{++}^D$, $\mathbf u_{0,F}$
is the unique optimizer of the following LP:
\begin{equation}
\boxed{
\begin{aligned}
    \underset{\mathbf u}{\min}
    \quad&
    \mathbf w^{\mathsf T}\mathbf u
    \\[1mm]
    \text{s.t.}\quad&
    \mathbf u\geq\mathbf f,
    \\
    &
    (\mathbf I-\mathbf K)\mathbf u\geq\mathbf a,
    \\
    &
    \mathbf0\leq\mathbf u\leq\mathbf1.
\end{aligned}}
\label{eq:direct_completion_LP}
\end{equation}
\end{corollary}

\begin{proof}
For the finite-dimensional operator in
\eqref{eq:finite_affine_operator}, the supersolution condition is
\[
    \mathbf T(\mathbf u)
    \leq
    \mathbf u.
\]
Using the affine form,
\[
    \mathbf a+\mathbf K\mathbf u
    \leq
    \mathbf u,
\]
which is equivalent to
\begin{equation}
    (\mathbf I-\mathbf K)\mathbf u
    \geq
    \mathbf a.
    \label{eq:finite_supersolution_linear}
\end{equation}
Hence $\mathcal S_{\mathbf f}$ is precisely the set of supersolution
vectors that dominate the analytical-bound vector $\mathbf f$.

The set is nonempty because $\mathbf1\geq\mathbf f$ and
$\mathbf T(\mathbf1)\leq\mathbf1$. The operator $\mathbf T$ is
monotone and maps $[0,1]^D$ into itself, so the sequence in
\eqref{eq:finite_completion_limit} is increasing and bounded.
Its limit therefore exists; continuity of the affine operator gives
$\mathbf T(\mathbf u_{0,F})\leq\mathbf u_{0,F}$, and
$\mathbf u_{0,F}\geq\mathbf f$. Thus
$\mathbf u_{0,F}\in\mathcal S_{\mathbf f}$. The leastness argument
of Proposition~\ref{prop:supersolution_completion} applies to the
same vector iteration and gives
\[
    \mathbf u_{0,F}
    \leq
    \mathbf u,
    \qquad
    \forall\,\mathbf u\in
    \mathcal S_{\mathbf f},
\]
which proves \eqref{eq:finite_componentwise_least}.

Now let $\mathbf w\in\mathbb R_{++}^D$ be arbitrary.  For every
$\mathbf u\in\mathcal S_{\mathbf f}$,
\[
    \mathbf u-\mathbf u_{0,F}
    \geq
    \mathbf0.
\]
Therefore,
\begin{equation}
\begin{aligned}
    \mathbf w^{\mathsf T}\mathbf u
    -
    \mathbf w^{\mathsf T}\mathbf u_{0,F}
    &=
    \mathbf w^{\mathsf T}
    \bigl(
        \mathbf u-\mathbf u_{0,F}
    \bigr)
    \\
    &=
    \sum_{i=1}^{D}
        w_i
        \bigl(
            u_i-u_{0,F,i}
        \bigr)
    \geq0.
\end{aligned}
\label{eq:positive_weight_scalarization}
\end{equation}
Since every $w_i>0$, equality in
\eqref{eq:positive_weight_scalarization}
holds only when
$\mathbf u=\mathbf u_{0,F}$.
Hence $\mathbf u_{0,F}$ is the unique optimizer of
\eqref{eq:direct_completion_LP}.
\end{proof}

Because $\mathbf u_{0,F}$ is the componentwise least element of
$\mathcal S_{\mathbf f}$, every positive weight vector $\mathbf w$
(in particular $\mathbf w=\mathbf1$) yields the same unique optimizer,
directly computing the completion limit in \eqref{eq:finite_completion_limit}.

Proposition~\ref{prop:supersolution_completion} guarantees validity and
minimality of the full-domain completion $\mathcal C(F)$, but does not
alone exclude $U_0\equiv1$. By \eqref{eq:T_one_fixed},
such a completed envelope would remain uninformative under further
operator iteration. We next give a sufficient
condition on the analytical input that rules out this degeneration
without requiring uniform tail decay over the state space.
Let $\pi$ denote the stationary law of $Z_k$. For $0<\varepsilon<1$, define
the state-dependent level threshold
\begin{equation}
 \ell_\varepsilon(z)
 :=\min\{m\in\mathbb N_0:F(m,z)\leq\varepsilon\}.
\label{eq:input_level_threshold}
\end{equation}
The auxiliary cutoff uses an integer grid in a fixed workload unit.
By monotonicity of $F$,
$x\geq\ell_\varepsilon(z)$ implies $F(x,z)\leq\varepsilon$.
The pointwise tail condition \eqref{eq:analytical_bound_tail} makes
$\ell_\varepsilon(z)$ finite, while the additional regularity condition is
\begin{equation}
 \int_{\mathsf Z}\ell_\varepsilon(z)\,\pi(dz)
 =\sum_{m=0}^{\infty}
   \pi\{z:F(m,z)>\varepsilon\}<\infty.
\label{eq:input_level_integrability}
\end{equation}
The equality is the tail-sum identity, since
$\{\ell_\varepsilon>m\}=\{F(m,\cdot)>\varepsilon\}$.
Thus the condition controls the stationary mean threshold needed to lower
the input bound to a fixed probability level.

For the tail and convergence results, assume that the stationary
net-input increment sequence is ergodic, $\mathbb E|Y_k|<\infty$, and
\begin{equation}
    \bar y\triangleq\mathbb E[Y_k]<0.
\label{eq:negative_mean_net_input}
\end{equation}
On the stationary backward marked path, set $\widetilde S_0=0$ and
$\widetilde S_m:=\sum_{\ell=1}^{m}Y_{k-\ell}$ for $m\geq1$.
By Birkhoff's ergodic theorem~\cite{birkhoff1931proof}, the
following limit holds almost surely (a.s.):
\begin{equation}
    \frac{\widetilde S_m}{m}
    =\frac1m\sum_{\ell=1}^{m}Y_{k-\ell}
    \longrightarrow\bar y
    \qquad\text{a.s.}
\label{eq:backward_net_input_ergodic}
\end{equation}
For a present threshold $x\geq0$, define the backward query thresholds
\begin{equation}
    \tilde x_0=x,\qquad \tilde x_m=x-\widetilde S_m,\qquad m\geq1,
\label{eq:backward_threshold_process}
\end{equation}
and the continuation events $A_0=\Omega$ and
\begin{equation}
    A_n:=\bigcap_{\ell=1}^{n}
    \{R_{k-\ell}\leq \tilde x_{\ell-1}\},\qquad n\geq1.
\label{eq:backward_continuation_event}
\end{equation}
The event $A_n$ means that none of the most recent $n$ blocks has already
generated the current violation independently of earlier workload.
On $A_n$, every queried threshold is nonnegative:
$\tilde x_\ell\geq R_{k-\ell}-Y_{k-\ell}\geq0$ for $1\leq\ell\leq n$.

\begin{proposition}[Nondegeneracy and Tail Preservation of Completion]
\label{prop:completion_nondegeneracy}
Assume the stationary ergodicity, integrability, and negative drift used
in \eqref{eq:backward_net_input_ergodic}. Let $F$ be an analytical workload
bound and $U_0=\mathcal C(F)$.
Assume that there exists $\varepsilon\in(0,1)$ such that
\eqref{eq:input_level_integrability} holds. Then
\begin{equation}
 \limsup_{x\to\infty}U_0(x,z)\leq\varepsilon<1.
\label{eq:completion_nondegenerate_tail}
\end{equation}
In particular, $U_0$ is not identically one.
If the condition holds for a sequence $\varepsilon_j\downarrow0$, then
\begin{equation}
 \lim_{x\to\infty}U_0(x,z)=0.
\label{eq:completion_statewise_proper}
\end{equation}
\end{proposition}

\begin{proof}
Fix such an $\varepsilon$ and set
$L_\varepsilon:=\ell_\varepsilon(Z_0)$. By stationarity and Tonelli's
theorem, for any $\eta>0$,
\[
\begin{aligned}
 &\sum_{m\geq1}
   \mathbb P\bigl(\ell_\varepsilon(Z_{k-m})>\eta m\bigr)\\
 &\quad=\mathbb E\!\left[
       \sum_{m\geq1}\mathbf1_{\{\eta m<L_\varepsilon\}}\right]\\
 &\quad\leq\frac{\mathbb E[L_\varepsilon]}{\eta}<\infty.
\end{aligned}
\]
The first Borel--Cantelli lemma~\cite{williams1991probability} therefore
gives $\ell_\varepsilon(Z_{k-m})\leq\eta m$ eventually a.s.
Taking $\eta=1/j$ for $j\geq1$ gives
\begin{equation}
 \frac{\ell_\varepsilon(Z_{k-m})}{m}\longrightarrow0
 \qquad\text{a.s.}
\label{eq:input_threshold_sublinear}
\end{equation}
Together with \eqref{eq:backward_net_input_ergodic}, this yields
$(\widetilde S_m+\ell_\varepsilon(Z_{k-m}))/m\to\bar y<0$, and hence
$\widetilde S_m+\ell_\varepsilon(Z_{k-m})\to-\infty$ a.s.

On the same stationary path, define the auxiliary process
\begin{equation}
 V_k^\varepsilon
 :=\sup_{m\geq0}\left\{
 \sum_{\ell=1}^{m}Y_{k-\ell}
       +\ell_\varepsilon(Z_{k-m})\right\}<\infty
 \quad\text{a.s.},
\label{eq:completion_level_remote_maximum}
\end{equation}
where the empty sum is zero. For $m\geq1$, separating the common
increment $Y_{k-1}$ and setting $r=m-1$ gives
\[
\begin{aligned}
 &\sup_{m\geq1}\left\{
     \sum_{\ell=1}^{m}Y_{k-\ell}
       +\ell_\varepsilon(Z_{k-m})\right\}\\
 &\quad=Y_{k-1}+\sup_{r\geq0}\left\{
     \sum_{j=1}^{r}Y_{k-1-j}
       +\ell_\varepsilon(Z_{k-1-r})\right\}\\
 &\quad=Y_{k-1}+V_{k-1}^\varepsilon.
\end{aligned}
\]
Combining this with the $m=0$ term $\ell_\varepsilon(Z_k)$ yields
\[
 V_k^\varepsilon
 =\max\{\ell_\varepsilon(Z_k),
             V_{k-1}^\varepsilon+Y_{k-1}\}.
\]
The stationary comparison workload
$\widehat W_k^\varepsilon:=\max\{W_k,V_k^\varepsilon\}$ is finite and,
by \eqref{eq:reflected_block_recursion}, satisfies
\begin{equation}
 \widehat W_k^\varepsilon
 =\max\bigl\{\widehat W_{k-1}^\varepsilon+Y_{k-1},\,
 R_{k-1},\,\ell_\varepsilon(Z_k)\bigr\}.
\label{eq:completion_auxiliary_workload_recursion}
\end{equation}
Write $\widehat Q_\varepsilon(x,z):=
\mathbb P(\widehat W_k^\varepsilon>x\mid Z_k=z)$.
The incoming comparison workload $\widehat W_{k-1}^\varepsilon$
is $\mathcal F_{k-1}^-$-measurable. By
\eqref{eq:marked_predictive_kernel}, conditioning additionally on
$(Y_{k-1},R_{k-1},Z_k)$ given $Z_{k-1}$ does not change its
conditional law. Stationarity identifies its tail
with $\widehat Q_\varepsilon(\cdot,Z_{k-1})$. Splitting the violation
event into $\{R_{k-1}>x\}$ and
$\{R_{k-1}\leq x,\,\widehat W_{k-1}^\varepsilon>x-Y_{k-1}\}$,
as in \eqref{eq:block_violation_decomposition}, therefore gives
\begin{equation}
\begin{aligned}
 &(\mathcal T\widehat Q_\varepsilon)(x,z)\\
 &\quad=\mathbb P\Bigl(
    \max\{\widehat W_{k-1}^\varepsilon+Y_{k-1},R_{k-1}\}>x
    \,\Big|\, Z_k=z\Bigr)\\
 &\quad\leq\mathbb P(\widehat W_k^\varepsilon>x\mid Z_k=z)
    =\widehat Q_\varepsilon(x,z),
\end{aligned}
\label{eq:completion_auxiliary_supersolution}
\end{equation}
where the inequality follows pathwise from
\eqref{eq:completion_auxiliary_workload_recursion}.
Almost-sure finiteness also gives
$\widehat Q_\varepsilon(x,z)\to0$ as $x\to\infty$.

Define
\begin{equation}
 U_\varepsilon
 :=\varepsilon\mathbf1+(1-\varepsilon)\widehat Q_\varepsilon.
\label{eq:completion_comparison_envelope}
\end{equation}
If $F(x,z)>\varepsilon$, then $x<\ell_\varepsilon(z)$, so
\eqref{eq:completion_auxiliary_workload_recursion} gives
$\widehat Q_\varepsilon(x,z)=1$ and $U_\varepsilon(x,z)=1$.
Otherwise, $U_\varepsilon(x,z)\geq\varepsilon\geq F(x,z)$.
Hence $U_\varepsilon\geq F$. Since $\mathcal T$ is affine,
\eqref{eq:T_one_fixed} and
\eqref{eq:completion_auxiliary_supersolution} imply
\[
\begin{aligned}
 \mathcal TU_\varepsilon
 &=\varepsilon\mathcal T\mathbf1
       +(1-\varepsilon)\mathcal T\widehat Q_\varepsilon\\
 &\leq\varepsilon\mathbf1
       +(1-\varepsilon)\widehat Q_\varepsilon
 =U_\varepsilon.
\end{aligned}
\]
The leastness of completion therefore gives $U_0\leq U_\varepsilon$.
Letting $x\to\infty$ proves
\eqref{eq:completion_nondegenerate_tail}. For a sequence
$\varepsilon_j\downarrow0$, take the intersection of the corresponding
state sets and let $j\to\infty$ to obtain
\eqref{eq:completion_statewise_proper}.

\end{proof}

For a complete predictive state with finite stationary support
$\mathsf Z_\pi=\{z_1,\ldots,z_M\}$, write $\pi_i=\pi(\{z_i\})>0$.
The defining tail condition \eqref{eq:analytical_bound_tail} already
makes every $\ell_\varepsilon(z_i)$ finite. Hence, for each
$0<\varepsilon<1$,
\begin{equation}
\begin{aligned}
 \int_{\mathsf Z}\ell_\varepsilon(z)\,\pi(dz)
 &=\sum_{i=1}^{M}\pi_i\ell_\varepsilon(z_i)\\
 &\leq\max_{1\leq i\leq M}\ell_\varepsilon(z_i)<\infty.
\end{aligned}
\label{eq:finite_state_level_integrability}
\end{equation}
Under the assumptions of Proposition~\ref{prop:completion_nondegeneracy},
it follows that $\max_{z\in\mathsf Z_\pi}U_0(x,z)\to0$ as $x\to\infty$.
For an infinite predictive-state space, however, pointwise tail decay
alone does not ensure \eqref{eq:input_level_integrability}; this
condition requires separate verification when applying
Proposition~\ref{prop:completion_nondegeneracy}.

\subsection{Monotone Queue-Dynamic Refinement}
\label{subsec:queue_dynamic_refinement}

Let $U_0$ denote a probability-valued workload envelope satisfying
\begin{equation}
    Q\leq U_0,
    \qquad
    \mathcal TU_0\leq U_0.
\label{eq:admissible_initial_supersolution}
\end{equation}
Under the analytical-completion construction of
Subsection~\ref{subsec:supersolution}, such an initial supersolution is
obtained as
\[
    U_0
    =
    \mathcal C(F).
\]

Starting from $U_0$, define
\begin{equation}
    U_{n+1}
    =
    \mathcal TU_n,
    \qquad
    n\geq0.
\label{eq:queue_dynamic_refinement}
\end{equation}

\begin{proposition}[Monotone Queue-Dynamic Refinement]
\label{prop:certified_queue_dynamic_refinement}
For every $n\geq0$,
\begin{equation}
    Q
    \leq
    U_{n+1}
    \leq
    U_n
    \leq
    U_0.
\label{eq:certified_hierarchy_order}
\end{equation}
Moreover,
\begin{equation}
    U_n-Q
    =
    \mathcal K^n(U_0-Q).
\label{eq:exact_refinement_error}
\end{equation}
\end{proposition}

\begin{proof}
Since $U_0$ is a supersolution,
\[
    U_1
    =
    \mathcal TU_0
    \leq
    U_0.
\]
Since $Q\leq U_0$, monotonicity of $\mathcal T$ and
$Q=\mathcal TQ$ give
\[
    Q
    =
    \mathcal TQ
    \leq
    \mathcal TU_0
    =
    U_1.
\]
Applying the same argument inductively yields
\eqref{eq:certified_hierarchy_order}.

Using
$\mathcal Tf=a+\mathcal Kf$
and
$Q=a+\mathcal KQ$,
\begin{equation}
    U_{n+1}-Q
    =
    \mathcal K(U_n-Q).
    \label{eq:one_step_refinement_error}
\end{equation}
Iterating \eqref{eq:one_step_refinement_error} gives
\eqref{eq:exact_refinement_error}.
\end{proof}

Proposition~\ref{prop:certified_queue_dynamic_refinement}
establishes the certified monotone ordering at every finite refinement
depth.  Convergence to the exact stationary workload tail requires, in
addition, that the residual dependence on the initial envelope vanish as
the refinement depth increases.  To characterize this mechanism, define the
initial discrepancy
\begin{equation}
    \Delta_0(x,z)
    \triangleq
    U_0(x,z)-Q(x,z)\geq0.
    \label{eq:initial_refinement_discrepancy}
\end{equation}

Use the backward thresholds and continuation events defined in
\eqref{eq:backward_threshold_process} and
\eqref{eq:backward_continuation_event}. On $A_n$, dependence on earlier
workload reaches the pair $(\tilde x_n,Z_{k-n})$.

At each backward step, the conditional law of the earlier mark and state
given the blocks already revealed is \eqref{eq:reverse_marked_kernel}.
Thus iterated conditional expectation expresses $\mathcal K^n$ as the
continuation expectation along this path, and
\eqref{eq:exact_refinement_error} gives
\begin{equation}
\begin{aligned}
    U_n(x,z)-Q(x,z)
    =
    \mathbb E\Big[
        \mathbf1_{A_n}
        \Delta_0(\tilde x_n,Z_{k-n})
        \,\Big|\,Z_k=z
    \Big].
\end{aligned}
\label{eq:refinement_error_pathwise}
\end{equation}

Equation~\eqref{eq:refinement_error_pathwise} gives a direct interpretation
of the refinement error.  A discrepancy in the initial envelope can affect
the present threshold after $n$ refinement steps only along paths for which
the workload dependence survives all $n$ intervening blocks, namely on
$A_n$.  Along such a surviving path, the remaining discrepancy is evaluated
at the backward threshold $\tilde x_n$.

The backward drift relation \eqref{eq:backward_net_input_ergodic} gives
\begin{equation}
    \frac{\tilde x_m}{m}
    =
    \frac{x-\widetilde S_m}{m}
    \longrightarrow
    -\bar y
    >
    0
    \qquad\text{a.s.},
\label{eq:backward_threshold_linear_escape}
\end{equation}
and therefore
\begin{equation}
    \tilde x_m
    \longrightarrow
    +\infty
    \qquad\text{a.s.}
\label{eq:backward_threshold_escape}
\end{equation}

Thus, although a surviving dependence path may extend arbitrarily far into the past, the corresponding past workload must lie in an increasingly extreme upper-tail region in order to affect the present exceedance event.  In
particular, for every finite $M$,
\begin{equation}
    \mathbb P\!\left(
        A_m,\,
        \tilde x_m\leq M
        \mid Z_k=z
    \right)
    \longrightarrow0.
    \label{eq:remote_threshold_escape}
\end{equation}

\begin{corollary}[Convergence under Remote-Threshold Escape]
\label{cor:remote_threshold_escape_convergence}
Suppose that the assumptions leading to
\eqref{eq:backward_threshold_escape} hold and that $U_0$ satisfies
\eqref{eq:admissible_initial_supersolution}. Assume either that $U_0$
is uniformly proper,
\begin{equation}
    \sup_{z\in\mathsf Z_*}U_0(x,z)\longrightarrow0,
    \qquad x\to\infty,
\label{eq:proper_initial_supersolution}
\end{equation}
or that $U_0=\mathcal C(F)$ and
\eqref{eq:input_level_integrability} holds for a sequence
$\varepsilon_j\downarrow0$. Then
\begin{equation}
    U_n(x,z)\downarrow Q(x,z).
\label{eq:queue_dynamic_convergence}
\end{equation}
For $U_0=\mathcal C(F)$ with an analytical workload bound $F$ and finite
$\mathsf Z_\pi$, the second alternative follows from
\eqref{eq:finite_state_level_integrability}.
\end{corollary}

\begin{proof}
Under \eqref{eq:proper_initial_supersolution}, splitting the expectation
in \eqref{eq:refinement_error_pathwise} at any finite $M>0$ gives
\begin{equation}
\begin{aligned}
 0\leq U_n(x,z)-Q(x,z)
 &\leq\mathbb P(A_n,\tilde x_n\leq M\mid Z_k=z)\\
 &\quad+\sup_{\substack{y>M\\s\in\mathsf Z_*}}U_0(y,s).
\end{aligned}
\label{eq:remote_threshold_error_bound}
\end{equation}
The first term vanishes as $n\to\infty$ by
\eqref{eq:remote_threshold_escape}; the second vanishes as $M\to\infty$.

For the second alternative, fix a retained level $\varepsilon_j$ and
write it as $\varepsilon$. Define the auxiliary-workload tail
$q_\varepsilon(x,z):=\mathbb P(V_k^\varepsilon>x\mid Z_k=z)$, with
$V_k^\varepsilon$ as in \eqref{eq:completion_level_remote_maximum}.
Since $\widehat W_k^\varepsilon=\max\{W_k,V_k^\varepsilon\}$,
splitting the exceedance event gives
\[
\begin{aligned}
 &0\leq\widehat Q_\varepsilon(x,z)-Q(x,z)\\
 &\quad=\mathbb P(W_k\leq x,\,V_k^\varepsilon>x\mid Z_k=z)\\
 &\quad\leq q_\varepsilon(x,z).
\end{aligned}
\]
The proof of Proposition~\ref{prop:completion_nondegeneracy} gives
$U_0\leq U_\varepsilon$. Using
\eqref{eq:completion_comparison_envelope} and then substituting the
preceding bound on $\widehat Q_\varepsilon-Q$ gives, at each $(x,z)$
with the common arguments suppressed,
\begin{equation}
\begin{aligned}
 0\leq U_0-Q
 &\leq U_\varepsilon-Q\\
 &=\varepsilon(1-Q)
      +(1-\varepsilon)(\widehat Q_\varepsilon-Q)\\
 &\leq\varepsilon(1-Q)+(1-\varepsilon)q_\varepsilon\\
 &\leq\varepsilon+q_\varepsilon.
\end{aligned}
\label{eq:completion_level_tail_bound}
\end{equation}
Positivity of $\mathcal K$,
$\mathcal K^n\mathbf1\leq\mathbf1$, and
\eqref{eq:exact_refinement_error} imply
\[
 0\leq U_n(x,z)-Q(x,z)
 \leq\varepsilon+(\mathcal K^nq_\varepsilon)(x,z).
\]
The backward path representation of $\mathcal K^n$ gives
\[
\begin{aligned}
 (\mathcal K^nq_\varepsilon)(x,z)
 =\mathbb P\bigl(&A_n,\,
 V_{k-n}^\varepsilon>\tilde x_n
 \mid Z_k=z\bigr).
\end{aligned}
\]
By the definition of the auxiliary process,
\[
 \widetilde S_n+V_{k-n}^\varepsilon
 =\sup_{m\geq n}\{\widetilde S_m+\ell_\varepsilon(Z_{k-m})\}.
\]
Since $\tilde x_n=x-\widetilde S_n$, dropping $A_n$ therefore gives
\begin{equation}
\begin{aligned}
 &U_n(x,z)-Q(x,z)\\
 &\quad\leq\varepsilon+
 \mathbb P\!\left(
 \sup_{m\geq n}\{\widetilde S_m+\ell_\varepsilon(Z_{k-m})\}>x
 \,\middle|\,Z_k=z\right).
\end{aligned}
\label{eq:regular_input_refinement_error}
\end{equation}
The proof of Proposition~\ref{prop:completion_nondegeneracy} gives
$\widetilde S_m+\ell_\varepsilon(Z_{k-m})\to-\infty$ a.s. Thus the tail
supremum tends to $-\infty$, and the conditional probability tends to zero.
First let $n\to\infty$ and then let $\varepsilon_j\downarrow0$ on the
common state set for the retained levels to obtain $U_n(x,z)-Q(x,z)\to0$.
The monotone decrease follows from
Proposition~\ref{prop:certified_queue_dynamic_refinement}.
\end{proof}

\begin{corollary}[Continuous Selection of Certified Bounds]
\label{cor:continuous_certified_refinement}
Let $\{U_n\}$ be the hierarchy in
Proposition~\ref{prop:certified_queue_dynamic_refinement}. For a real-valued
refinement index $t\geq0$, set $n=\lfloor t\rfloor$, $\delta=t-n$, and
$U(t)=(1-\delta)U_n+\delta U_{n+1}$.
The family $U(t)\geq Q$ is continuous and nonincreasing in $t$ at each
state--threshold pair. For every integer $k\geq0$,
\begin{equation}
\begin{aligned}
 \mathcal T^k U(t)&=U(t+k)\leq U(t),\\
 U(t+k)-Q&=\mathcal K^k\bigl(U(t)-Q\bigr).
\end{aligned}
\label{eq:continuous_refinement_shift}
\end{equation}
Thus every $U(t)$ is a supersolution. Under the conditions of
Corollary~\ref{cor:remote_threshold_escape_convergence}, $U(t)\downarrow Q$
as $t\to\infty$.
\end{corollary}

\begin{proof}
The linear segments join at the integer levels, and
$Q\leq U_{n+1}\leq U(t)\leq U_n$ gives the ordering and the limit.
Affineness of $\mathcal T^k$ sends the two endpoints to $U_{n+k}$ and
$U_{n+k+1}$ with unchanged weights. Since $\mathcal TQ=Q$,
subtracting $Q$ from successive updates gives the error identity.
\end{proof}

Starting from $F_0^{\mathrm{clip}}=F$, the pointwise clipping map
$\Psi(f)\triangleq f\wedge\mathcal Tf$ gives
$F_{n+1}^{\mathrm{clip}}=\Psi(F_n^{\mathrm{clip}})$.
Order preservation and $F\leq U_0$ give
$Q\leq F_{n+1}^{\mathrm{clip}}\leq F_n^{\mathrm{clip}}\leq U_n$,
so $F_n^{\mathrm{clip}}\downarrow Q$ whenever $U_n\downarrow Q$.
The error update includes a nonlinear correction:
$F_{n+1}^{\mathrm{clip}}-Q=\mathcal K(F_n^{\mathrm{clip}}-Q)
-[\mathcal T F_n^{\mathrm{clip}}-F_n^{\mathrm{clip}}]^+$.
For the piecewise-linear interpolant $F^{\mathrm{clip}}(t)$ of the clipped
iterates, the concavity of $\Psi$ gives
$\Psi(F^{\mathrm{clip}}(t))\geq F^{\mathrm{clip}}(t+1)$, and the inequality
can be strict. Thus clipping and interpolation need not commute.
In contrast, Corollary~\ref{cor:continuous_certified_refinement} gives a
continuous family of supersolutions $U(t)$ that tighten directly under
$\mathcal T$, with exact shifts and linear error propagation.

In summary, Theorem~\ref{thm:exact_workload_fixed_point} and
Lemma~\ref{lem:T_order_properties} supply the exact affine operator.
Propositions~\ref{prop:supersolution_completion}
and~\ref{prop:direct_completion_limit} identify the least supersolution
dominating $F$, and Corollary~\ref{cor:direct_finite_completion} gives its
LP computation for an exact finite-dimensional realization.
Proposition~\ref{prop:completion_nondegeneracy} addresses nondegeneracy and
tail preservation. Proposition~\ref{prop:certified_queue_dynamic_refinement}
gives monotone refinement, with convergence under
Corollary~\ref{cor:remote_threshold_escape_convergence}, and
Corollary~\ref{cor:continuous_certified_refinement} extends the hierarchy
to continuous selection of supersolutions. When the LP realization
and the convergence conditions both apply, we obtain
\begin{equation}
    F
    \xrightarrow{\text{direct LP}}
    U_0
    \geq U_1
    \geq U_2
    \geq\cdots\downarrow Q.
\label{eq:complete_certified_hierarchy}
\end{equation}

\section{Palm Transfer from Workload Bounds to Packet Completion Delay}
\label{sec:palm_transfer}

Virtual delay $V(t)$ in \eqref{eq:virtual_delay} is time-sampled and depends
on future realizations of the random service process; serialization
corrections~\cite{leboudec2001network} account for packetization while
preserving that time-sampling law. To obtain packet-indexed completion-delay
guarantees from the closed affine workload dynamics in
Section~\ref{sec:workload_refinement}, we transfer the state-conditioned
workload hierarchy through an exact packet-Palm functional.
Stationary averaging and this transfer yield
unconditional workload and packet-completion DVP hierarchies, respectively,
preserving validity, monotonicity, and convergence.

\subsection{Packet-Palm Representation of Completion Delay}
\label{subsec:packet_palm_representation}

The workload analysis in Section~\ref{sec:workload_refinement} is formulated
at stationary queue-dynamic reference boundaries, whereas packet completion
delay is defined from the viewpoint of an arriving packet. We connect these
two viewpoints through the Palm distribution of the stationary packet-arrival
process.

Let $\mathbb P^{A}$ and $\mathbb E^{A}$ denote probability and expectation
under the packet-Palm distribution.  For a finite observation interval
$B$ with $0<\mathbb E[N(B)]<\infty$ and a nonnegative packet-centered
functional $G$, the Campbell--Palm identity~\cite{baccelli2003elements} gives
\begin{equation}
    \mathbb E^{A}[G]
    =
    \frac{
        \mathbb E\!\left[
            \sum_{T_i\in B} G_i
        \right]
    }{
        \mathbb E[N(B)]
    },
    \label{eq:packet_palm_identity}
\end{equation}
where $N(B)$ counts packets and $G_i$ evaluates $G$ with packet $i$ tagged.
A batch of $N_b$ packets contributes $N_b$ terms. Choosing
$G_i=\mathbf1_{\{D_i>d\}}$ gives the completion-delay tail of a typical
packet.

For block-based sampling take $B=(0,H]$ and let $N_0=N(B)$ be the block
packet count. Let $W_0$ and $Z_0$ be the workload and predictive state
at the start of the tagged packet's block.
Under the ordinary stationary measure, the conditional workload tail is
\begin{equation}
    Q(x,z)
    \triangleq
    \mathbb P(W_0>x\mid Z_0=z),
    \qquad x\geq0.
    \label{eq:palm_boundary_workload_tail}
\end{equation}
Hence $Q$ is a deterministic function of the workload threshold and the
boundary state.

Under $\mathbb P^{A}$, let $\tau\in(0,H]$ be the arrival phase of the
tagged batch and $r$ the tagged packet's FIFO rank. Write $L_{\leq r}$
for its prefix workload, omitting the batch index. Let $Y_\tau$ and
$R_\tau$ be the net increment and the restart workload along the
partial block up to $\tau^-$, excluding the whole tagged batch even
when $\tau=H$. The workload immediately before the batch satisfies
\begin{equation}
    W_{\tau^-}
    =
    \max\{W_0+Y_\tau,R_\tau\}.
\label{eq:tagged_prearrival_workload}
\end{equation}
The term $W_0+Y_\tau$ represents workload carried from the reference
boundary through the partial path, whereas $R_\tau$ represents workload
generated after a possible emptying of the queue.

For a fixed delay threshold $d>0$, define
\begin{equation}
    S_d^+
    \triangleq
    S(\tau,\tau+d]
    \label{eq:tagged_future_service}
\end{equation}
as the cumulative service available from the tagged arrival until its
deadline.  The quantity $S_d^+$ is a random variable determined by the
realized service process over $(\tau,\tau+d]$.

Let $D_{\rm pkt}$ denote the completion delay of the tagged packet.
Under FIFO, the pre-batch workload and the prefix through the tagged
packet must be served by $\tau+d$ to meet its deadline, so
\begin{equation}
    D_{\rm pkt}>d
    \quad\Longleftrightarrow\quad
    W_{\tau^-}+L_{\leq r}>S_d^+.
\label{eq:packet_completion_event}
\end{equation}

Substituting \eqref{eq:tagged_prearrival_workload} into
\eqref{eq:packet_completion_event} gives
\[
    D_{\rm pkt}>d
    \quad\Longleftrightarrow\quad
    \max\{W_0+Y_\tau,R_\tau\}+L_{\leq r}>S_d^+.
\]
Separating the two mutually exclusive cases
$R_\tau+L_{\leq r}>S_d^+$ and $R_\tau+L_{\leq r}\leq S_d^+$ yields
\begin{equation}
\begin{aligned}
    \mathbf1_{\{D_{\rm pkt}>d\}}
    ={}&
    \mathbf1_{\{R_\tau+L_{\leq r}>S_d^+\}}
    \\
    &+
    \mathbf1_{\{R_\tau+L_{\leq r}\leq S_d^+\}}
    \mathbf1_{\{
        W_0>S_d^+-Y_\tau-L_{\leq r}
    \}}.
\end{aligned}
\label{eq:packet_completion_decomposition}
\end{equation}
The first term represents a violation already implied by the reflected
post-boundary evolution and the tagged batch prefix. On its
complement, the delay violation is determined by whether the workload
present at the reference boundary exceeds the remaining service margin.

Define this remaining margin as the random workload threshold
\begin{equation}
    \Xi_d
    \triangleq
    S_d^+-Y_\tau-L_{\leq r}.
\label{eq:palm_random_workload_threshold}
\end{equation}
With this threshold, \eqref{eq:packet_completion_decomposition} becomes
\begin{equation}
\begin{aligned}
    \mathbf1_{\{D_{\rm pkt}>d\}}
    ={}&
    \mathbf1_{\{R_\tau+L_{\leq r}>S_d^+\}}
    \\
    &+
    \mathbf1_{\{R_\tau+L_{\leq r}\leq S_d^+\}}
    \mathbf1_{\{W_0>\Xi_d\}}.
\end{aligned}
\label{eq:packet_completion_decomposition_xi}
\end{equation}

The remaining task is therefore to express the conditional probability
of the event $\{W_0>\Xi_d\}$ under the packet-Palm measure in terms of the
stationary workload tail $Q$ characterized in
Section~\ref{sec:workload_refinement}. This requires two conditional
compatibility properties.

First, we impose Palm state compatibility: packet-Palm sampling
preserves the workload distribution within a fixed boundary state,
\begin{equation}
    \mathbb P^{A}
    \left(
        W_0>x
        \,\middle|\,
        Z_0
    \right)
    =
    \mathbb P
    \left(
        W_0>x
        \,\middle|\,
        Z_0
    \right)
    =
    Q(x,Z_0),
    \qquad x\geq0.
\label{eq:palm_state_compatibility}
\end{equation}
Packet-Palm sampling weights the boundary-state law by the conditional
mean packet count per block. By \eqref{eq:packet_palm_identity},
\eqref{eq:palm_state_compatibility} holds if $N_0$ and $W_0$ are
conditionally independent given $Z_0$.

Second, define the tagged-context sigma-field
\begin{equation}
    \mathcal G_d
    \triangleq
    \sigma\!\left(
        Z_0,
        \tau,
        L_{\leq r},
        Y_\tau,
        R_\tau,
        S_d^+
    \right).
    \label{eq:palm_tagged_context_sigmafield}
\end{equation}
We impose the corresponding boundary-separation condition
\begin{equation}
    \mathbb P^{A}
    \left(
        W_0>x
        \,\middle|\,
        \mathcal G_d
    \right)
    =
    \mathbb P^{A}
    \left(
        W_0>x
        \,\middle|\,
        Z_0
    \right),
    \qquad x\geq0,
\label{eq:palm_boundary_separation}
\end{equation}
a.s. In the block-count construction, both Palm conditions follow if,
under $\mathbb P$, $W_0$ is conditionally independent of the packet count
and the collection of tagged contexts in $(0,H]$ given $Z_0$. Indeed,
this independence factors $Q(x,Z_0)$ out of the conditional expectation
of each packet-weighted context sum in \eqref{eq:packet_palm_identity}.

Combining \eqref{eq:palm_state_compatibility} and
\eqref{eq:palm_boundary_separation} gives the workload bridge
\begin{equation}
\begin{aligned}
    \mathbb P^{A}
    \left(
        W_0>x
        \,\middle|\,
        \mathcal G_d
    \right)
    &=
    \mathbb P^{A}
    \left(
        W_0>x
        \,\middle|\,
        Z_0
    \right)
    \\
    &=
    \mathbb P
    \left(
        W_0>x
        \,\middle|\,
        Z_0
    \right)
    \\
    &=
    Q(x,Z_0),
    \qquad x\geq0 .
\end{aligned}
\label{eq:palm_boundary_workload_bridge}
\end{equation}

The relation in \eqref{eq:palm_boundary_workload_bridge} is an equality
of conditional workload-tail functions. On the continuation event
$\{R_\tau+L_{\leq r}\leq S_d^+\}$, the random threshold in
\eqref{eq:palm_random_workload_threshold} satisfies
$\Xi_d\geq R_\tau-Y_\tau\geq0$. Since $\Xi_d$ is
$\mathcal G_d$-measurable, this conditional tail can be evaluated at
$\Xi_d$. On this event,
\begin{equation}
    \mathbb E^{A}\!\left[
        \mathbf1_{\{W_0>\Xi_d\}}
        \,\middle|\,
        \mathcal G_d
    \right]
    =
    Q(\Xi_d,Z_0).
\label{eq:palm_random_threshold_bridge}
\end{equation}
Here $Q$ is a fixed deterministic function, evaluated at the random pair
$(\Xi_d,Z_0)$ on the continuation event.

Equation~\eqref{eq:palm_random_threshold_bridge} provides the key
connection between the workload analysis and the tagged-packet event in
\eqref{eq:packet_completion_decomposition_xi}.  On the branch
$R_\tau+L_{\leq r}>S_d^+$, the delay violation is already determined by the
post-boundary evolution.  On the complementary branch, its conditional
probability is obtained by evaluating the stationary workload tail at
the random pair $(\Xi_d,Z_0)$.

To express this connection in a form that applies both to the exact
workload tail $Q$ and to the certified workload bounds developed in
Section~\ref{sec:workload_refinement}, let
$f:[0,\infty)\times\mathsf Z\rightarrow[0,1]$
be any deterministic measurable probability-valued workload-tail function.
Its evaluation $f(\Xi_d,Z_0)$ is used only on
$\{R_\tau+L_{\leq r}\leq S_d^+\}$, where the threshold is nonnegative.
Define the packet-Palm transfer functional
\begin{equation}
\begin{aligned}
    \mathcal H_d(f)
    \triangleq
    \mathbb E^{A}\Big[
        &\mathbf1_{\{R_\tau+L_{\leq r}>S_d^+\}}
        \\
        &+
        \mathbf1_{\{R_\tau+L_{\leq r}\leq S_d^+\}}
        f(\Xi_d,Z_0)
    \Big].
\end{aligned}
\label{eq:palm_transfer_functional}
\end{equation}
The functional $\mathcal H_d$ maps a
boundary-workload tail function into the corresponding packet-level
violation probability generated by the tagged-packet sampling law.

\begin{lemma}[Exact Packet-Palm Transfer]
\label{lem:exact_packet_palm_transfer}
Under the Palm state-compatibility condition
\eqref{eq:palm_state_compatibility} and the boundary-separation condition
\eqref{eq:palm_boundary_separation}, the stationary
packet-completion DVP satisfies
\begin{equation}
    p_{\rm pkt}(d)
    \triangleq
    \mathbb P^{A}(D_{\rm pkt}>d)
    =
    \mathcal H_d(Q).
\label{eq:exact_packet_palm_transfer}
\end{equation}
\end{lemma}

\begin{proof}
Define
\begin{equation}
    A_d
    \triangleq
    \{R_\tau+L_{\leq r}>S_d^+\}.
    \label{eq:palm_local_violation_event}
\end{equation}
Its complement is $A_d^c=\{R_\tau+L_{\leq r}\leq S_d^+\}$.
Then \eqref{eq:packet_completion_decomposition_xi} becomes
\begin{equation}
    \mathbf1_{\{D_{\rm pkt}>d\}}
    =
    \mathbf1_{A_d}
    +
    \mathbf1_{A_d^c}
    \mathbf1_{\{W_0>\Xi_d\}}.
    \label{eq:palm_indicator_decomposition_compact}
\end{equation}
Taking packet-Palm expectation gives
\begin{equation}
\begin{aligned}
    p_{\rm pkt}(d)
    ={}&
    \mathbb E^{A}[\mathbf1_{A_d}]
    \\
    &+
    \mathbb E^{A}\!\left[
        \mathbf1_{A_d^c}
        \mathbf1_{\{W_0>\Xi_d\}}
    \right].
\end{aligned}
\label{eq:palm_transfer_proof_step1}
\end{equation}

The event $A_d$ depends only on
$R_\tau$, $L_{\leq r}$, and $S_d^+$, and hence
$A_d\in\mathcal G_d$.  Likewise,
$\Xi_d=S_d^+-Y_\tau-L_{\leq r}$ is
$\mathcal G_d$-measurable.  Applying the tower property to the second
term in \eqref{eq:palm_transfer_proof_step1} therefore gives
\begin{equation}
\begin{aligned}
&
\mathbb E^{A}\!\left[
    \mathbf1_{A_d^c}
    \mathbf1_{\{W_0>\Xi_d\}}
\right]
\\
&\qquad=
\mathbb E^{A}\!\left[
    \mathbb E^{A}\!\left[
        \mathbf1_{A_d^c}
        \mathbf1_{\{W_0>\Xi_d\}}
        \,\middle|\,
        \mathcal G_d
    \right]
\right]
\\
&\qquad=
\mathbb E^{A}\!\left[
    \mathbf1_{A_d^c}
    \mathbb E^{A}\!\left[
        \mathbf1_{\{W_0>\Xi_d\}}
        \,\middle|\,
        \mathcal G_d
    \right]
\right].
\end{aligned}
\label{eq:palm_transfer_tower}
\end{equation}
Using \eqref{eq:palm_random_threshold_bridge},
\begin{equation}
\begin{aligned}
&
\mathbb E^{A}\!\left[
    \mathbf1_{A_d^c}
    \mathbf1_{\{W_0>\Xi_d\}}
\right]
\\
&\qquad=
\mathbb E^{A}\!\left[
    \mathbf1_{A_d^c}
    Q(\Xi_d,Z_0)
\right].
\end{aligned}
\label{eq:palm_transfer_carry_term}
\end{equation}
Substituting this identity into
\eqref{eq:palm_transfer_proof_step1} yields
\begin{equation}
\begin{aligned}
    p_{\rm pkt}(d)
    =
    \mathbb E^{A}\Big[
        &\mathbf1_{\{R_\tau+L_{\leq r}>S_d^+\}}
        \\
        &+
        \mathbf1_{\{R_\tau+L_{\leq r}\leq S_d^+\}}
        Q(\Xi_d,Z_0)
    \Big].
\end{aligned}
\label{eq:palm_transfer_Q_expanded}
\end{equation}
By \eqref{eq:palm_transfer_functional}, the right-hand side is exactly
$\mathcal H_d(Q)$, which proves
\eqref{eq:exact_packet_palm_transfer}.
\end{proof}

For later use, the packet-Palm transfer functional admits the affine
decomposition
\begin{equation}
    \mathcal H_d(f)
    =
    b_d+\mathcal P_df,
\label{eq:palm_affine_decomposition}
\end{equation}
where
\begin{equation}
\begin{aligned}
    b_d
    &\triangleq
    \mathbb P^{A}(R_\tau+L_{\leq r}>S_d^+),
    \\
    \mathcal P_df
    &\triangleq
    \mathbb E^{A}\!\left[
        \mathbf1_{\{R_\tau+L_{\leq r}\leq S_d^+\}}
        f(\Xi_d,Z_0)
    \right].
\end{aligned}
\label{eq:palm_linear_component}
\end{equation}
The mapping $\mathcal P_d$ is a positive linear functional of $f$.
The term $b_d$ accounts for violations already determined by the
post-boundary reflected evolution, whereas $\mathcal P_df$ accounts for
the part that remains dependent on the reference-boundary workload tail.

\subsection{Unconditional Workload and Packet-Delay Hierarchies}
\label{subsec:palm_hierarchy_transfer}

We now pass from the state-conditioned profiles used in completion and
refinement to unconditional performance guarantees. Let
$\pi(dz)=\mathbb P(Z_0\in dz)$ be the ordinary stationary law at a
queue-dynamic reference boundary. For a workload profile $f$, define
\begin{equation}
\begin{aligned}
 \bar f(x)&:=\int_{\mathsf Z}f(x,z)\,\pi(dz),\\
 \bar Q(x)&=\mathbb P(W_0>x)
          =\int_{\mathsf Z}Q(x,z)\,\pi(dz).
\end{aligned}
\label{eq:stationary_workload_state_average}
\end{equation}
Thus $\bar U_n$ bounds the unconditional reference-boundary workload tail.

For packet delay, use instead the boundary-state law seen by a typical
packet, $\nu^{A}(dz):=\mathbb P^{A}(Z_0\in dz)$. With $A_d$ from
\eqref{eq:palm_local_violation_event}, define
\begin{equation}
\begin{aligned}
 \mathcal H_{d,z}(f):=
 \mathbb E^{A}\!\left[
  \mathbf1_{A_d}+\mathbf1_{A_d^c}f(\Xi_d,z)
  \,\middle|\,Z_0=z\right].
\end{aligned}
\label{eq:conditional_packet_palm_functional}
\end{equation}
Conditional expectation then disintegrates the Palm functional as
\begin{equation}
 \mathcal H_d(f)
 =\int_{\mathsf Z}\mathcal H_{d,z}(f)\,\nu^{A}(dz).
\label{eq:packet_palm_state_disintegration}
\end{equation}
For a fixed queue configuration, $\pi$ and $\nu^{A}$ are fixed by the
queue and its packet-sampling rule, independently of refinement depth.
The integral representation applies to finite, continuous, and mixed
predictive-state spaces.

For block-count sampling, set $m(z):=\mathbb E[N_0\mid Z_0=z]$ and
$\bar m:=\mathbb E[N_0]\in(0,\infty)$.
Applying \eqref{eq:packet_palm_identity} to a bounded nonnegative measurable
boundary-state function $\varphi$ gives
$\mathbb E^{A}[\varphi(Z_0)]
=\mathbb E[N_0\varphi(Z_0)]/\bar m$. Hence
\begin{equation}
 \nu^{A}(dz)
 =\frac{m(z)}{\bar m}\,\pi(dz),
 \qquad
 \bar m=\int_{\mathsf Z}m(z)\,\pi(dz).
\label{eq:packet_palm_boundary_state_law}
\end{equation}
For a block-constant conditional packet intensity $\lambda(z)$,
$m(z)=H\lambda(z)$. The density $m(z)/\bar m$ gives $\nu^{A}\ll\pi$,
with $\nu^{A}=\pi$ exactly when $m(z)=\bar m$ for $\pi$-almost every $z$.

For any two measurable probability-valued workload-tail functions
$f,g:[0,\infty)\times\mathsf Z\rightarrow[0,1]$ satisfying
\[
    f(x,z)\leq g(x,z)
    \qquad
    \text{for all }(x,z),
\]
the positive linearity of $\mathcal P_d$ in
\eqref{eq:palm_affine_decomposition} gives
\begin{equation}
    f\leq g
    \quad\Longrightarrow\quad
    \mathcal H_d(f)
    \leq
    \mathcal H_d(g).
\label{eq:palm_transfer_monotonicity}
\end{equation}
Thus, an upper ordering between workload-tail functions is preserved after
packet-Palm transfer.

The transfer functional also preserves monotone limits.  Suppose
\[
    f_n(x,z)\downarrow f(x,z)
\]
pointwise, with $0\leq f_n\leq1$. On $A_d^c$,
$(\Xi_d,Z_0)\in[0,\infty)\times\mathsf Z$, so
\[
    f_n(\Xi_d,Z_0)
    \downarrow
    f(\Xi_d,Z_0)
\]
a.s. on $A_d^c$. Hence the corresponding integrands in
\eqref{eq:palm_transfer_functional} decrease pointwise and remain bounded
between zero and one.  By the bounded convergence theorem,
\[
    \mathcal H_d(f_n)
    \longrightarrow
    \mathcal H_d(f).
\]
Together with \eqref{eq:palm_transfer_monotonicity}, this yields
\begin{equation}
    f_n\downarrow f
    \quad\Longrightarrow\quad
    \mathcal H_d(f_n)
    \downarrow
    \mathcal H_d(f).
\label{eq:palm_transfer_continuity}
\end{equation}

\begin{proposition}[From State-Conditioned to Unconditional Hierarchies]
\label{prop:palm_preservation_hierarchy}

Assume the Palm compatibility and boundary separation of
Lemma~\ref{lem:exact_packet_palm_transfer}. Starting from
$U_0=\mathcal C(F)$, let $\{U_n\}_{n\geq0}$ be the certified hierarchy
of Subsection~\ref{subsec:queue_dynamic_refinement},
write $\bar U_n$ and $\bar F$ for the averages in
\eqref{eq:stationary_workload_state_average}, and define
\begin{equation}
    B_n(d)
    \triangleq
    \mathcal H_d(U_n)
    =\int_{\mathsf Z}\mathcal H_{d,z}(U_n)\,\nu^{A}(dz).
\label{eq:packet_delay_bound_hierarchy}
\end{equation}
Then, for every finite refinement depth $n$,
\begin{equation}
 \bar Q(x)\leq\bar U_{n+1}(x)\leq\bar U_n(x),\qquad x\geq0,
\label{eq:unconditional_workload_hierarchy_order}
\end{equation}
\begin{equation}
    p_{\rm pkt}(d)
    \leq
    B_{n+1}(d)
    \leq
    B_n(d).
\label{eq:packet_delay_hierarchy_order}
\end{equation}
The analytical input and its completion also satisfy
\begin{equation}
\begin{aligned}
 \bar Q(x)&\leq\bar F(x)\leq\bar U_0(x),\\
 p_{\rm pkt}(d)&\leq\mathcal H_d(F)\leq B_0(d).
\end{aligned}
\label{eq:unconditional_analytical_completion_order}
\end{equation}

With $\Delta_0$ as defined in \eqref{eq:initial_refinement_discrepancy},
the workload error at depth $n$ is
\begin{equation}
 \bar U_n(x)-\bar Q(x)
 =\int_{\mathsf Z}(\mathcal K^n\Delta_0)(x,z)\,\pi(dz),
\label{eq:unconditional_workload_refinement_error}
\end{equation}
and the packet-delay error is
\begin{align}
 B_n(d)-p_{\rm pkt}(d)
 &=\mathcal P_d(U_n-Q),\label{eq:palm_error_linear}\\
 &=\mathcal P_d\mathcal K^n\Delta_0.\label{eq:packet_delay_refinement_error_compact}
\end{align}

If, for almost every state $z$ under both $\pi$ and $\nu^{A}$,
the convergence in Corollary~\ref{cor:remote_threshold_escape_convergence}
holds simultaneously for all $x\geq0$, then
\begin{equation}
 \bar U_n(x)\downarrow\bar Q(x),\qquad x\geq0,
\label{eq:unconditional_workload_hierarchy_convergence}
\end{equation}
\begin{equation}
    B_n(d)
    \downarrow
    p_{\rm pkt}(d).
\label{eq:packet_delay_hierarchy_convergence}
\end{equation}
\end{proposition}

\begin{proof}
Integrating $Q\leq U_{n+1}\leq U_n$ under $\pi$ proves
\eqref{eq:unconditional_workload_hierarchy_order}.
The same statewise ordering, \eqref{eq:palm_transfer_monotonicity},
and $\mathcal H_d(Q)=p_{\rm pkt}(d)$ from
Lemma~\ref{lem:exact_packet_palm_transfer} prove
\eqref{eq:packet_delay_hierarchy_order}. Applying these positive
mappings to $Q\leq F\leq U_0$ gives
\eqref{eq:unconditional_analytical_completion_order}.
Averaging \eqref{eq:exact_refinement_error} under $\pi$ gives
\eqref{eq:unconditional_workload_refinement_error}.
In $\mathcal H_d(U_n)-\mathcal H_d(Q)$, the common term $b_d$ in
\eqref{eq:palm_affine_decomposition} cancels, giving
\eqref{eq:palm_error_linear}. Substituting
\eqref{eq:exact_refinement_error} then gives
\eqref{eq:packet_delay_refinement_error_compact}. By
\eqref{eq:palm_linear_component}, the remaining difference is
\[
 \mathcal P_d(U_n-Q)
 =\mathbb E^A\!\left[
   \mathbf1_{A_d^c}(U_n-Q)(\Xi_d,Z_0)
 \right].
\]

For each fixed $x\geq0$, the integrand in
\eqref{eq:unconditional_workload_refinement_error} equals
$U_n(x,z)-Q(x,z)$, which is bounded by one and tends to zero for
$\pi$-almost every $z$. Applying the dominated convergence
theorem~\cite{williams1991probability} with dominating
function $1$ gives $\bar U_n(x)\to\bar Q(x)$.
Combining this limit with \eqref{eq:unconditional_workload_hierarchy_order}
proves \eqref{eq:unconditional_workload_hierarchy_convergence}.

For fixed $d>0$, $Z_0$ has law $\nu^A$ under $\mathbb P^A$.
Thus $U_n(x,Z_0)\to Q(x,Z_0)$ holds with $\mathbb P^A$-probability one
simultaneously for all $x\geq0$. On $A_d^c$,
\[
\begin{aligned}
\Xi_d
  &= S_d^+-Y_\tau-L_{\leq r}\\
  &= \underbrace{(S_d^+-R_\tau-L_{\leq r})}_{\geq 0\text{ on }A_d^c}
     +\underbrace{(R_\tau-Y_\tau)}_{\geq 0}
     \geq 0.
\end{aligned}
\]
We may therefore substitute $x=\Xi_d$ on $A_d^c$. This gives
$(U_n-Q)(\Xi_d,Z_0)\to0$ $\mathbb P^A$-a.s. on that event.
The Palm integrand is zero on $A_d$ and bounded by one, so bounded
convergence under $\mathbb P^A$ gives $B_n(d)\to p_{\rm pkt}(d)$.
The ordering in \eqref{eq:packet_delay_hierarchy_order} then gives
\eqref{eq:packet_delay_hierarchy_convergence}. In the block-count
construction, $\nu^A\ll\pi$ ensures that simultaneous convergence for
$\pi$-almost every state also holds for $\nu^A$-almost every state.
\end{proof}

The continuous levels in Corollary~\ref{cor:continuous_certified_refinement}
extend both unconditional hierarchies. Stationary averaging interpolates the
unconditional workload bounds. For packet delay, let
$B(t;d)=\mathcal H_d(U(t))$ with $n$ and $\delta$ as in that corollary.
By affineness of $\mathcal H_d$ and \eqref{eq:continuous_refinement_shift},
for every integer $k\geq0$,
\begin{equation}
\begin{aligned}
 B(t;d)&=(1-\delta)B_n(d)+\delta B_{n+1}(d),\\
 B(t+k;d)-p_{\rm pkt}(d)&=\mathcal P_d\mathcal K^k\bigl(U(t)-Q\bigr).
\end{aligned}
\label{eq:continuous_packet_refinement}
\end{equation}
For fixed $d$, $B(t;d)$ is a valid upper bound, nonincreasing in $t$,
and converges to $p_{\rm pkt}(d)$ under the convergence conditions of
Proposition~\ref{prop:palm_preservation_hierarchy}. Any target between
$B_{n+1}(d)$ and $B_n(d)$ is attained at a depth in $[n,n+1]$. The
selected workload envelope remains certified for further refinement and
defines a single DVP bound curve $B(t;\cdot)$ with the same $t$ at every
delay threshold.

\section{Queue-Model Instantiations and Numerical Validation}
\label{sec:numerical_validation}

We test four wireless queues: a GI/GI/1-type baseline, MMCPP
arrivals, AR($q$)-driven packetized traffic, and Poisson--Pareto
batch arrivals. They cover independent blocks, finite-state and
continuous-state dependence, and heavy-tailed bursts. All experiments
apply the state-conditioned completion and refinement in
\eqref{eq:completion_fixed_point} and \eqref{eq:queue_dynamic_refinement}.
The resulting unconditional workload and packet-delay hierarchies follow
Proposition~\ref{prop:palm_preservation_hierarchy}, illustrating the distinct
roles of state dependence, packetization, and heavy-tailed bursts.

\subsection{Common Wireless System and Evaluation Setup}
\label{subsec:common_wireless_setup}

The FIFO uplink has an unbounded buffer and independent arrival and
channel processes. Packet lengths are i.i.d. Let $L$ denote a generic
packet length in bits, so $L/8$ is uniform on $\{64,\ldots,1517\}$.
We set the transmission time interval (TTI) to $0.5$ ms.
Each $H=1$-ms fading block spans two TTIs. All traffic models have mean
arrival rate $2$ packets per TTI
($4000$ packets/s). Transmission is continuous and work-conserving:
unused service is discarded, and completion timestamps are recorded
directly in continuous time.
Delay includes waiting and transmission; workload counts all unfinished
bits, including the packet in service.

Table~\ref{tab:common_channel_states} specifies three propagation
states: line-of-sight (LOS), lower-loss non-line-of-sight (NLOS-L), and
higher-loss non-line-of-sight (NLOS-D). The GI/GI/1 baseline fixes the
NLOS-L propagation parameters, while the other three queues use
Markov-modulated (MM) wireless service, with a stationary Markov chain
governing the propagation state. The chain is updated every four TTIs ($2$ ms),
with $p_{ii}=0.8$ and $p_{ij}=0.1$ for $i\neq j$.
Small-scale fading follows a Nakagami-$m$ flat block-fading model~\cite{nakagami1960m}.
Conditional on the propagation state, the power gain satisfies
$G\sim\operatorname{Gamma}(m,1/m)$ with shape $m$ and scale $1/m$.
The gain remains constant within each $1$-ms block.

To compensate for state-dependent path loss, the transmitter uses
a simplified open-loop power-control model with full path-loss
compensation, motivated by New Radio (NR) uplink power control~\cite{3gpp38213}.
We set the target received power to $-75$ dBm and cap the transmit
power at $23$ dBm. With powers in dBm and path loss $PL(z)$ in dB,
the model sets $P_{\rm tx}(z)=\min\{23,-75+PL(z)\}$ and
$P_{\rm rx}(z)=P_{\rm tx}(z)-PL(z)$.
The bandwidth per resource block (RB) is $B_{\rm RB}=360$ kHz, and
$N_{\rm RB}$ denotes the number of allocated RBs.
Over an interval of duration $\Delta t$ with unchanged channel
realization, the available service is
\begin{equation}
 C(\Delta t)=N_{\rm RB}B_{\rm RB}\Delta t
 \log_2\!\left(1+
 \frac{\bar P_{\rm rx}(z)G}
 {N_{\rm RB}B_{\rm RB}N_0}\right).
\label{eq:common_wireless_service}
\end{equation}
Here $\bar P_{\rm rx}(z)$ denotes the received power in watts
corresponding to $P_{\rm rx}(z)$, and $N_0$ is the noise power
spectral density (PSD) in W/Hz, corresponding to $-174$ dBm/Hz.

\begin{table}[!htbp]
\centering
\caption{Wireless Propagation and Fading Parameters}
\label{tab:common_channel_states}
\begin{tabular}{lccc}
\toprule
State & $m$ & $PL$ (dB) & $P_{\rm rx}$ (dBm)\\
\midrule
GI/GI/1 (fixed) & 1.0 & 90  & $-75$\\
LOS             & 3.0 & 80  & $-75$\\
NLOS-L          & 1.0 & 90  & $-75$\\
NLOS-D          & 0.5 & 100 & $-77$\\
\bottomrule
\end{tabular}
\end{table}

Define the mean load as $\rho=\mathbb E[A_k(H)]/\mathbb E[S_k(H)]$.
To establish comparable baseline conditions across the four
queueing models, we choose RB allocations that yield mean loads
close to $0.7$. These baseline settings are used to evaluate
$U_0$ completion, the certified hierarchy, and refinement
convergence. Table~\ref{tab:common_resource_loads} summarizes
the baseline allocations and their actual mean loads, together
with additional RB configurations used to examine resource-dependent
refinement depth. Arrival parameters remain unchanged within
each resource comparison.

\begin{table}[!htbp]
\centering
\caption{Resource Allocations and Actual Mean Loads}
\label{tab:common_resource_loads}
\begin{tabular}{lcc}
\toprule
Scenario & $N_{\rm RB}$ & Mean load\\
\midrule
\multicolumn{3}{l}{Baseline configurations}\\
GI/GI/1               & 10 & 0.683522\\
MMCPP                 & 10 & 0.706406\\
AR($q$)-packetized    & 10 & 0.706406\\
Poisson--Pareto batch & 10 & 0.706406\\
\midrule
\multicolumn{3}{l}{Additional fixed allocations}\\
GI/GI/1               & 8  & 0.828533\\
MMCPP                 & 12 & 0.604449\\
Poisson--Pareto batch & 20 & 0.392030\\
\bottomrule
\end{tabular}
\end{table}

For each configuration, we simulate $300$ independent queue
realizations and collect $10^6$ packet completion delays from
each. At each threshold $d$, we compute the fraction of recorded
delays exceeding $d$ in each realization and average these fractions
to estimate the stationary packet-completion DVP.

Completion iterations are indexed by $r$, separately from refinement
depth $n$; completion comparisons are performed in the workload domain.
Unless a state--threshold slice is specified, refinement curves report
$\bar U_n(x)$ for workload and $B_n(d)$ for packet delay.

\subsection{GI/GI/1-Type Queue with Independent Block Fading}
\label{subsec:gigi1_validation}

We construct a GI/GI/1-type transmission queue by combining
homogeneous compound-Poisson packet arrivals with service over a
quasi-static flat Nakagami block-fading channel. We use the fixed
propagation parameters in Table~\ref{tab:common_channel_states}. Independent
arrivals and fading across $H=1$-ms blocks yield i.i.d. marked
blocks and a singleton predictive state. The arrival intensity is
$\lambda=2$ packets per TTI, giving a mean count of $2\lambda$ packets per
block. We write the packet-size, block-arrival, and block-service MGFs as
$M_L(\theta)=\mathbb E[e^{\theta L}]$,
$M_A(\theta)=\exp\{2\lambda[M_L(\theta)-1]\}$, and
$M_S(-\theta)=\mathbb E[e^{-\theta S_k(H)}]$, respectively,
and let $\psi(\theta)=M_A(\theta)M_S(-\theta)$.

Both analytical workload bounds follow by specializing
Appendix~\ref{app:markov_conditional_bounds} to a singleton predictive
state. Then $\check P=[1]$, $b_\theta=M_A(\theta)$, and
$G_\theta=\psi(\theta)$. The comparison $R_m\leq A_m$ accounts for
within-block restarts in both bounds. In the single-state case, the
resolvent factor in \eqref{eq:app_markov_resolvent_derivation} reduces to
$M_A(\theta)/(1-\psi(\theta))$ for $\psi(\theta)<1$. Taking the infimum
over admissible $\theta>0$ and capping at one yields the MGF-based SNC
bound~\cite{6868978} for $x\geq0$:
\begin{equation}
\begin{aligned}
 \mathbb P(W_0>x)&\leq F_{\rm C}(x),\\
 F_{\rm C}(x)&:=\min\!\left\{1,
 \inf_{\theta>0:\psi(\theta)<1}
 \frac{M_A(\theta)e^{-\theta x}}{1-\psi(\theta)}\right\}.
\end{aligned}
\label{eq:gigi1_analytical_bounds}
\end{equation}
For the exponential martingale bound, normalize $h_\theta=1$, so
$H_\theta=M_S(-\theta)$. At the positive root $\theta^\star$,
$\psi(\theta^\star)=1$ gives $1/H_{\theta^\star}=M_A(\theta^\star)$.
Evaluating
\eqref{eq:app_markov_conditional_stopping_bound} at $\theta^\star$ gives
\begin{equation}
\begin{aligned}
 \mathbb P(W_0>x)&\leq F_{\rm M}(x)
 :=\min\!\left\{1,\frac{e^{-\theta^\star x}}{M_S(-\theta^\star)}\right\},\\
 \theta^\star&=\sup\{\theta>0:\psi(\theta)\leq1\}.
\end{aligned}
\label{eq:gigi1_martingale_bound}
\end{equation}

\begin{figure}[!htbp]
\centering
\includegraphics[width=0.70\linewidth]{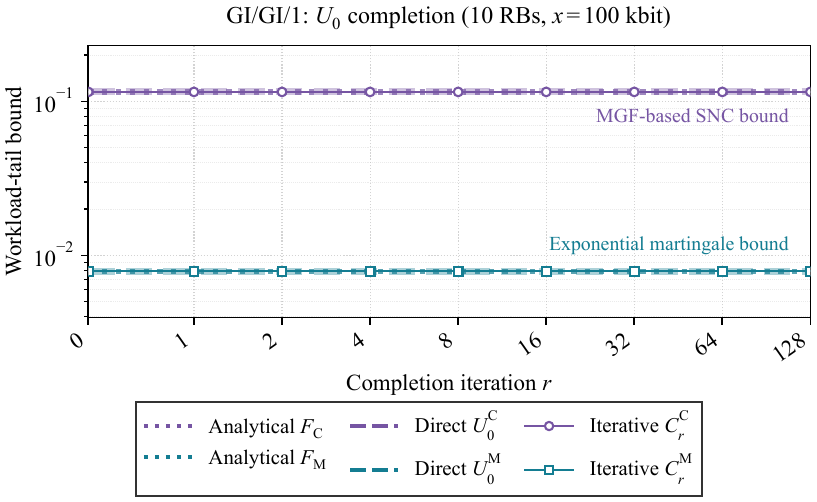}
\caption{Direct and iterative GI/GI/1 workload completion at $10$ RBs
and $x=100$ kbit for the MGF-based SNC and exponential martingale bounds.}
\label{fig:gigi1_completion}
\end{figure}

\paragraph{Single-state completion}
Figure~\ref{fig:gigi1_completion} compares the independent direct and
iterative workload completions at $10$ RBs and $x=100$ kbit.
For both starting bounds, the iterates $C_r$ through $r=128$ agree with
the direct solution $U_0$ at the plotted threshold, illustrating
Corollary~\ref{cor:direct_finite_completion}. Both completions also
coincide numerically with their corresponding analytical bounds there.

\begin{figure}[!htbp]
\centering
\includegraphics[width=\linewidth]{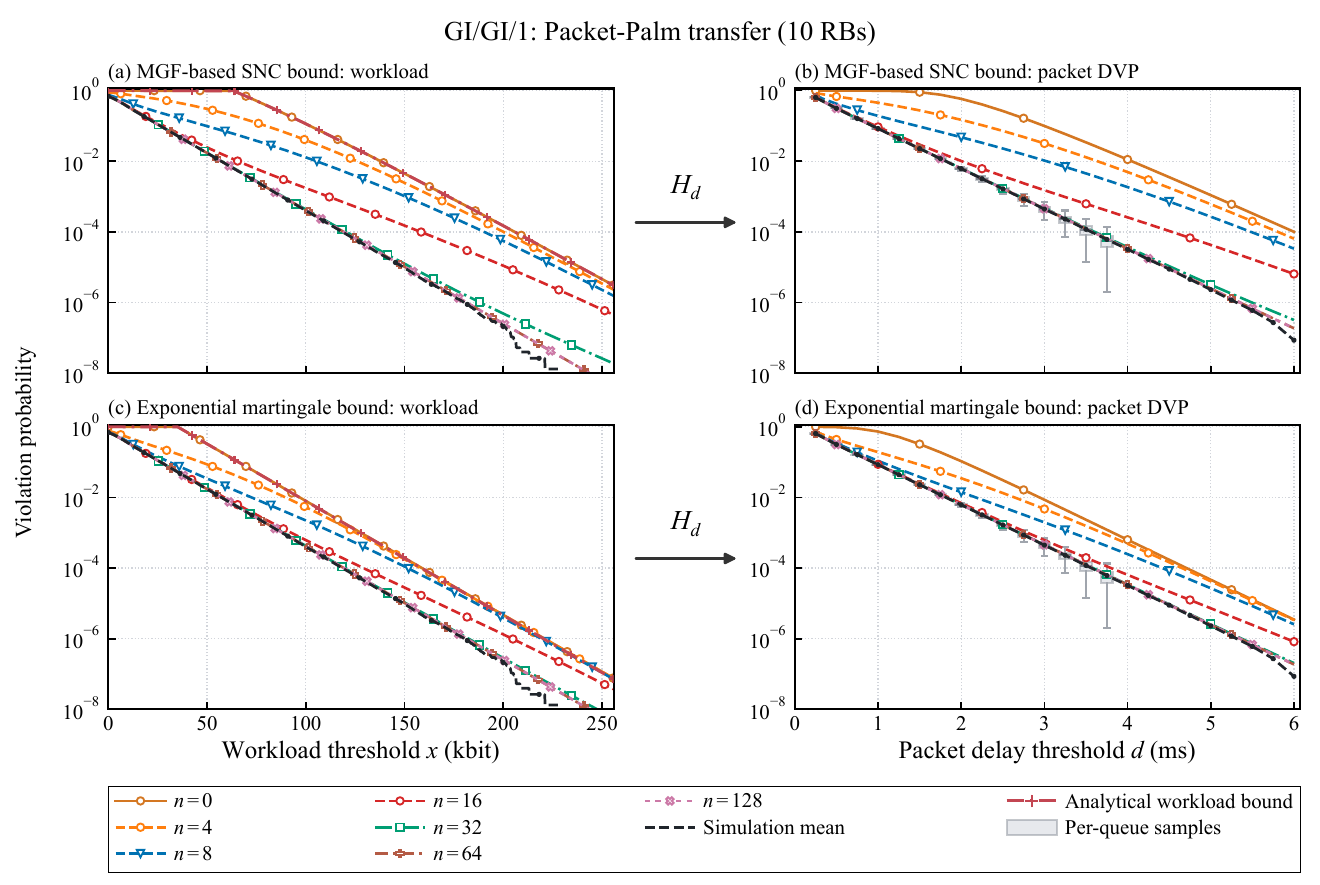}
\caption{GI/GI/1 workload hierarchies (a),(c) and their packet-DVP images
under Palm transfer (b),(d) at $10$ RBs through $n=128$. Panels (a),(b)
start from the MGF-based SNC bound and panels (c),(d) from the exponential
martingale bound.}
\label{fig:gigi1_palm}
\end{figure}

\paragraph{Workload refinement and Packet-Palm validation}
The Poisson input gives a uniform tagged phase and satisfies the
compatibility and separation conditions of
Lemma~\ref{lem:exact_packet_palm_transfer}.
With a singleton predictive state, the average in
\eqref{eq:stationary_workload_state_average} leaves the workload profiles
unchanged.
Figure~\ref{fig:gigi1_palm} shows workload hierarchies in panels (a),(c)
and their certified packet-DVP images in panels (b),(d), through $n=128$.
Refinement tightens both hierarchies, with a larger reduction from the
MGF-based SNC bound.
The workload reference samples unfinished bits at $1$-ms block boundaries.
Both hierarchies approach their respective empirical means, consistent with
the exact transfer and its packet-domain gap identity
\eqref{eq:palm_error_linear}.

\begin{figure}[!htbp]
\centering
\includegraphics[width=\linewidth]{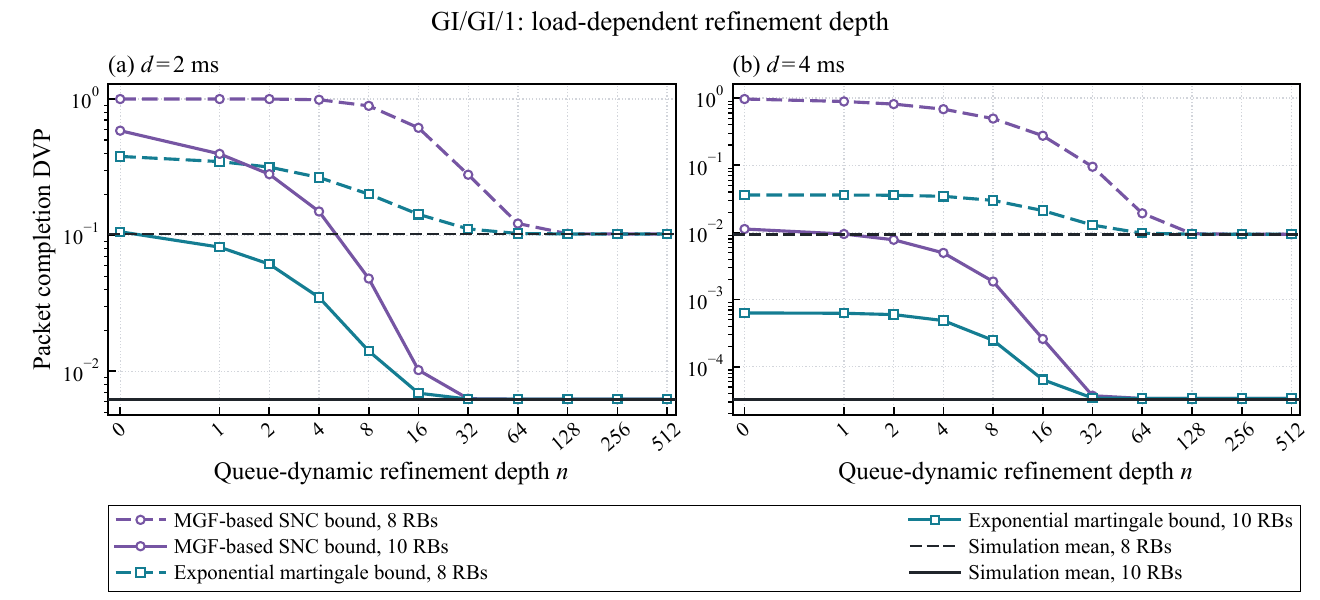}
\caption{GI/GI/1 packet-DVP refinement at $8$ and $10$ RBs through $n=512$,
with (a) $d=2$ ms and (b) $d=4$ ms.}
\label{fig:gigi1_load_convergence}
\end{figure}

\paragraph{Resource-dependent refinement depth}
In Fig.~\ref{fig:gigi1_load_convergence}, $8$ RBs require deeper
refinement than $10$ RBs from either starting bound. For example,
the MGF-based SNC curves are close to their empirical references by
$n=32$ at $10$ RBs. At $8$ RBs, a gap remains at $n=64$.
With the arrival process fixed, $10$ RBs provide a larger mean service
surplus per block. By \eqref{eq:backward_threshold_linear_escape}, this
increases the asymptotic growth rate of the backward query thresholds.
Faster threshold growth can move continuing queries sooner into the far
tail, where the initial gap $U_0-Q$ is small. This helps explain the
shallower refinement observed at $10$ RBs.
Equation~\eqref{eq:packet_delay_refinement_error_compact} then links the
remaining workload error to the packet DVP gap. Only the linear parts
of the two affine maps enter this calculation: $\mathcal K^n$ gives the
workload error after $n$ steps, and the Palm average $\mathcal P_d$
preserves its nonnegativity and monotone convergence to zero.

\subsection{MMCPP Arrival with MM Wireless Service}
\label{subsec:markov_validation}

The three-state arrival chain updates once per block. Within a block,
arrivals are Poisson with conditional intensities
$\boldsymbol\lambda=(0.8,2,8)$ packets per TTI, giving a mean count of
$2\lambda_i$ packets per block in state $i$. The arrival transition
matrix is
\begin{equation}
 P_A\simeq
 \begin{pmatrix}
 0.05&0.779388&0.170612\\
 0.926942&0.05&0.023058\\
 0.853061&0.096939&0.05
 \end{pmatrix}.
\label{eq:markov_arrival_transition}
\end{equation}
Its stationary law is $\pi_A=(0.49,0.412,0.098)$.
Each $2$-ms propagation update interval contains two fading blocks
of length $H$. The phase $\ell\in\{0,1\}$ identifies the first or second fading
block within that interval. Adding this phase to the three propagation
states gives a six-state service chain with transition matrix $P_S$.
The joint state is $z=(i,v)$, with arrival state $i$ and service state
$v$. The joint state space $\mathsf Z$ has $18$ elements. Independence of
arrivals and service gives $P=P_A\otimes P_S$ as the joint transition
matrix. Let $\check P$ denote the stationary reversal of
$P$~\cite{norris1997markov}.

\paragraph{State-conditioned analytical bounds}
We set the block transforms to
$a_i(\theta)=\exp\{2\lambda_i[M_L(\theta)-1]\}$ and
$s_v(\theta)=\mathbb E[e^{-\theta S_k(H)}\mid v]$.
For each $z=(i,v)$, the joint-state vectors $a_\theta$ and $s_\theta$
have entries $a_z(\theta)=a_i(\theta)$ and $s_z(\theta)=s_v(\theta)$,
respectively. The product $a_\theta s_\theta$ is componentwise.
With $G_\theta=\check P\operatorname{diag}(a_\theta s_\theta)$
and $b_\theta=\check P a_\theta$, we obtain the conditional
MGF-based SNC bound~\cite{bouillard2024mgf}:
\begin{equation}
\begin{aligned}
 &\mathbb P(W_0>x\mid Z_0=z)\leq F_{\rm C}(x,z),\\
 &F_{\rm C}(x,z):=\min\!\left\{1,
 \inf_{\theta>0:\operatorname{spr}(G_\theta)<1}
 [(I-G_\theta)^{-1}b_\theta]_z e^{-\theta x}\right\}.
\end{aligned}
\label{eq:markov_classical_bound}
\end{equation}
Here $\operatorname{spr}$ denotes spectral radius, and $I$ is the
identity matrix.

Using the stopping argument of~\cite{10.1145/3287322}, we also obtain a
conditional exponential martingale bound. Let $h_\theta>0$ be a right
Perron eigenvector of $G_\theta$. The exponent and bound are
\begin{equation}
\begin{aligned}
 \theta^\star&=\sup\{\theta>0:\operatorname{spr}(G_\theta)\leq1\},\\
 H_\theta&=\min_{j\in\mathsf Z}\{s_j(\theta)h_\theta(j)\},\\
 \mathbb P(W_0>x\mid Z_0=z)&\leq F_{\rm M}(x,z),\\
 F_{\rm M}(x,z)&:=\min\!\left\{1,
 \frac{h_{\theta^\star}(z)}{H_{\theta^\star}}
 e^{-\theta^\star x}\right\}.
\end{aligned}
\label{eq:markov_martingale_bound}
\end{equation}
The minimum in $H_\theta$ is over all joint states $j\in\mathsf Z$,
and $s_j(\theta)$ depends only on the service component of $j$.
Appendix~\ref{app:markov_conditional_bounds} proves both bounds.

\begin{figure*}[!htbp]
\centering
\includegraphics[width=\linewidth]{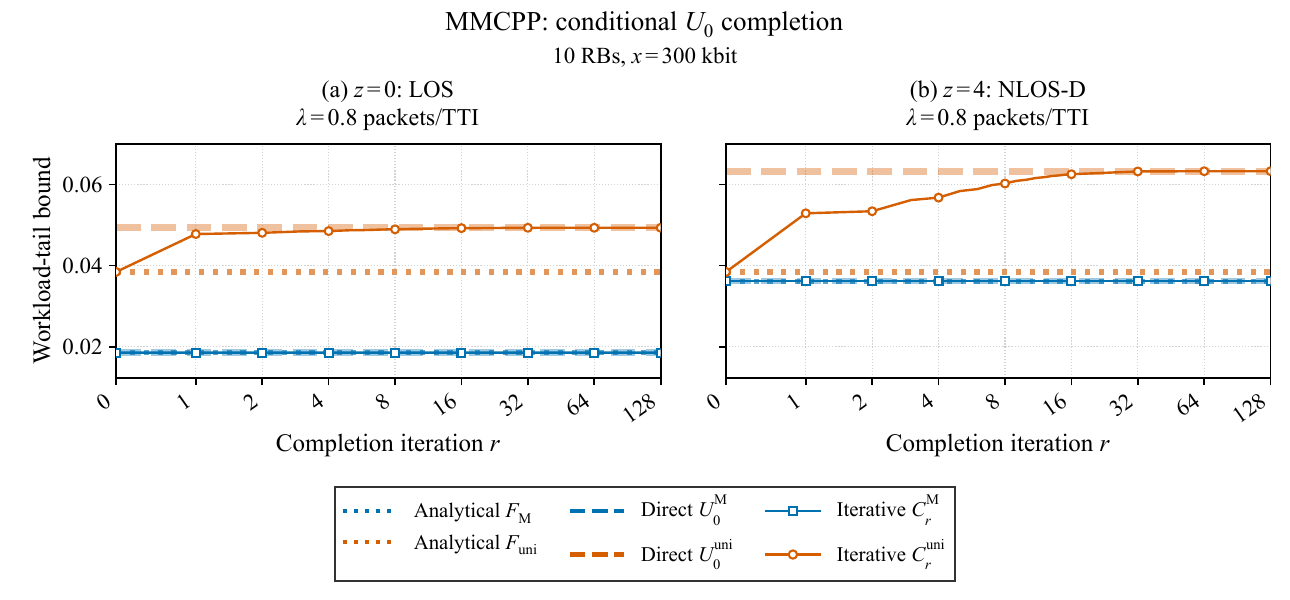}
\caption{Direct and iterative MMCPP workload completion for $F_{\rm M}$
and $F_{\rm uni}$ at $10$ RBs and $x=300$ kbit, with (a) $z=0$ (LOS)
and (b) $z=4$ (NLOS-D).}
\label{fig:markov_completion}
\end{figure*}

\begin{figure*}[!htbp]
\centering
\includegraphics[width=\linewidth]{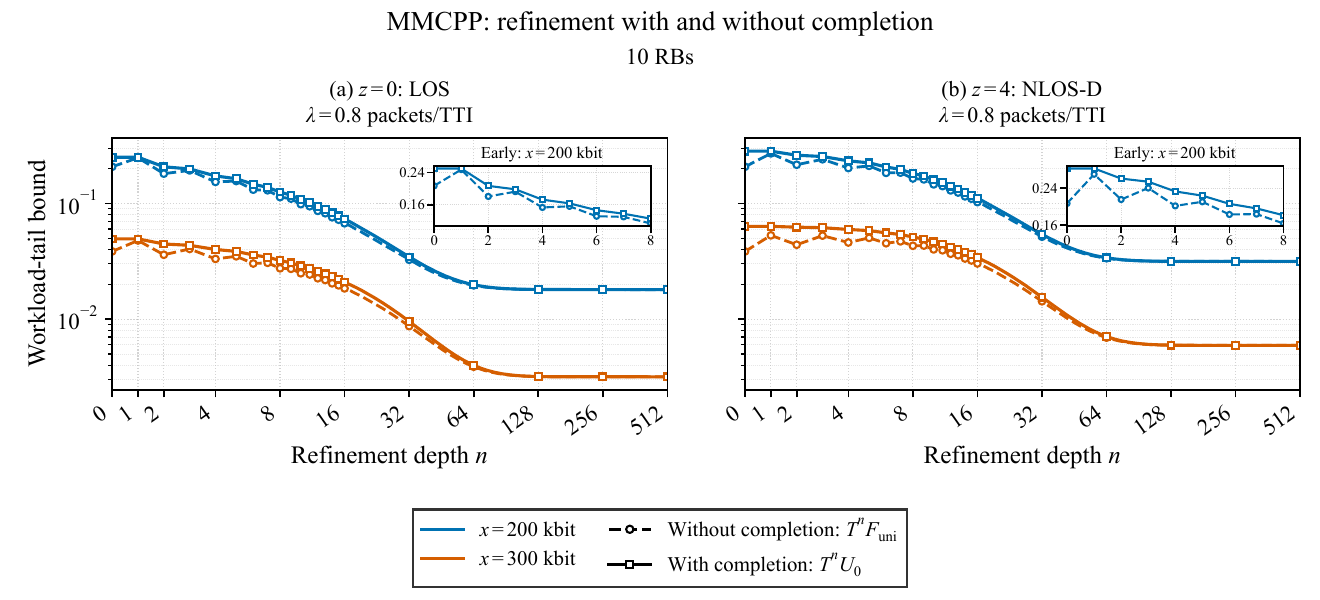}
\caption{MMCPP workload refinement from $F_{\rm uni}$ and its completion
$U_0$ at $10$ RBs through $n=512$. Both $x=200$ and $300$ kbit are shown
for (a) $z=0$ and (b) $z=4$.}
\label{fig:markov_completion_refinement}
\end{figure*}

\paragraph{Multistate completion and monotonicity}
We represent $F$ and every $U_n$ by $18$ state components on a common
workload grid. Completion jointly solves the finite-dimensional obstacle
problem in Corollary~\ref{cor:direct_finite_completion}, and refinement applies
\eqref{eq:exact_queue_operator} to the full state-dependent function.
Each update combines predecessor-state tails according to $\check P$.
Figures~\ref{fig:markov_completion} and~\ref{fig:markov_completion_refinement}
show individual state slices of the joint $18$-state computation.

To expose the role of completion, we also use the state-uniform envelope
\begin{equation}
 Q(x,z)\leq F_{\rm M}(x,z)\leq
 F_{\rm uni}(x,z):=\max_{z'\in\mathsf Z}F_{\rm M}(x,z').
\label{eq:markov_uniform_martingale_envelope}
\end{equation}
This defines a valid state-uniform analytical workload bound across all
states.

Figure~\ref{fig:markov_completion} compares both inputs at $10$ RBs and
$x=300$ kbit. States $z=0$ and $z=4$ share the low arrival intensity
$0.8$ packets per TTI and phase $\ell=0$, but correspond to
LOS and NLOS-D, respectively. Completion barely changes $F_{\rm M}$,
whereas the larger $F_{\rm uni}$ requires a visible lift at both states.
In each case, the displayed iterates $C_r$ approach the independently
computed direct $U_0$. The supersolution condition depends on the
propagated slack in \eqref{eq:global_slack_equivalence}.

Figure~\ref{fig:markov_completion_refinement} then compares
$\mathcal T^nF_{\rm uni}$ and $\mathcal T^nU_0$ at the same states for
$x=200$ and $300$ kbit. The direct iterates from $F_{\rm uni}$ show early
increases and decreases, and the two sequences nearly overlap at the
largest displayed depths. For the exact operator, order preservation
and $\mathcal TQ=Q$ ensure upper-bound validity for both sequences.
The supersolution property of $U_0$ ensures monotone tightening at
every depth for the completed sequence.

\paragraph{From conditional workload bounds to packet DVP}
For the packet-DVP experiments, we complete and refine the original
conditional bounds $F_{\rm C}$ and $F_{\rm M}$.
We then apply the Palm transfer to each refined hierarchy.
The stationary boundary law is $\pi_A(i)\pi_S(v)$, with $\pi_S$ the
six-state service-chain law. Here $m(i,v)=2\lambda_i$, so
\eqref{eq:packet_palm_state_disintegration} and
\eqref{eq:packet_palm_boundary_state_law} specialize to
\begin{equation}
\begin{aligned}
 B_n(d)&=\sum_{i,v}\omega_{i,v}\,
              \mathcal H_{d,(i,v)}(U_n),\\
 \omega_{i,v}&=
 \frac{\pi_A(i)\lambda_i\pi_S(v)}
      {\sum_j\pi_A(j)\lambda_j}.
\end{aligned}
\label{eq:markov_palm_state_mixture}
\end{equation}
The conditional functional is defined in
\eqref{eq:conditional_packet_palm_functional}.
Proposition~\ref{prop:palm_preservation_hierarchy} certifies the resulting
unconditional packet-DVP curves directly from the coupled $18$-state
workload hierarchy.

\begin{figure*}[!htbp]
\centering
\includegraphics[width=\linewidth]{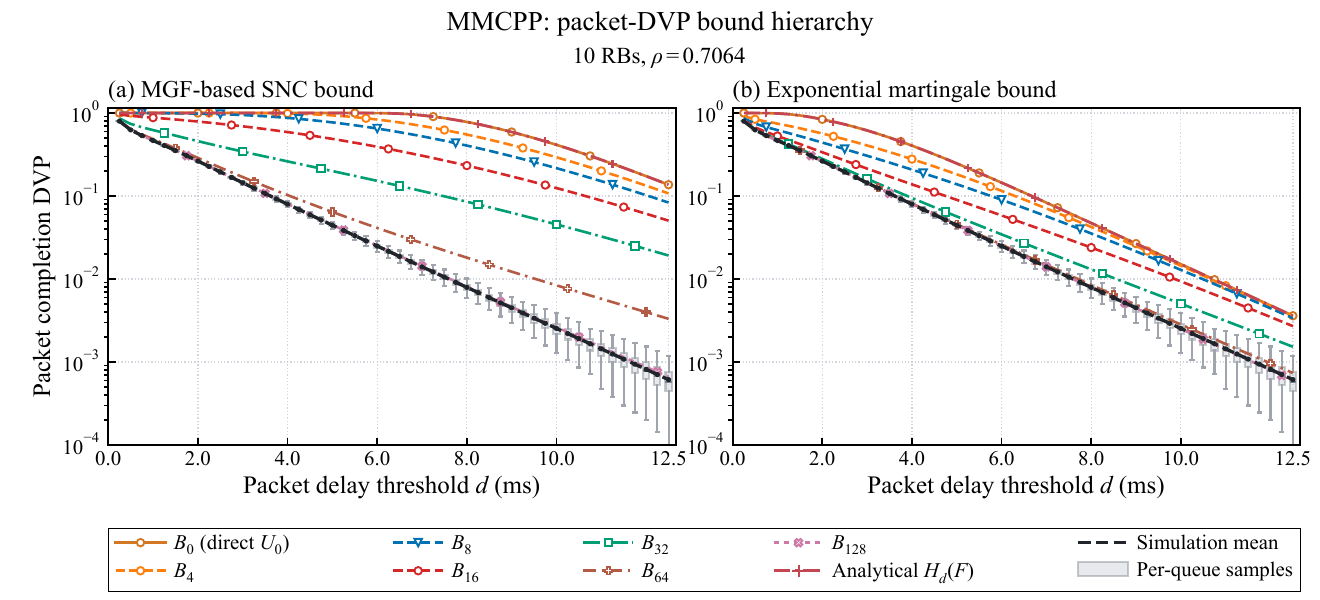}
\caption{MMCPP certified packet-DVP hierarchies at $10$ RBs through
$n=128$, starting from (a) the MGF-based SNC bound and (b) the
exponential martingale bound.}
\label{fig:markov_hierarchy}
\end{figure*}

\begin{figure*}[!htbp]
\centering
\includegraphics[width=\linewidth]{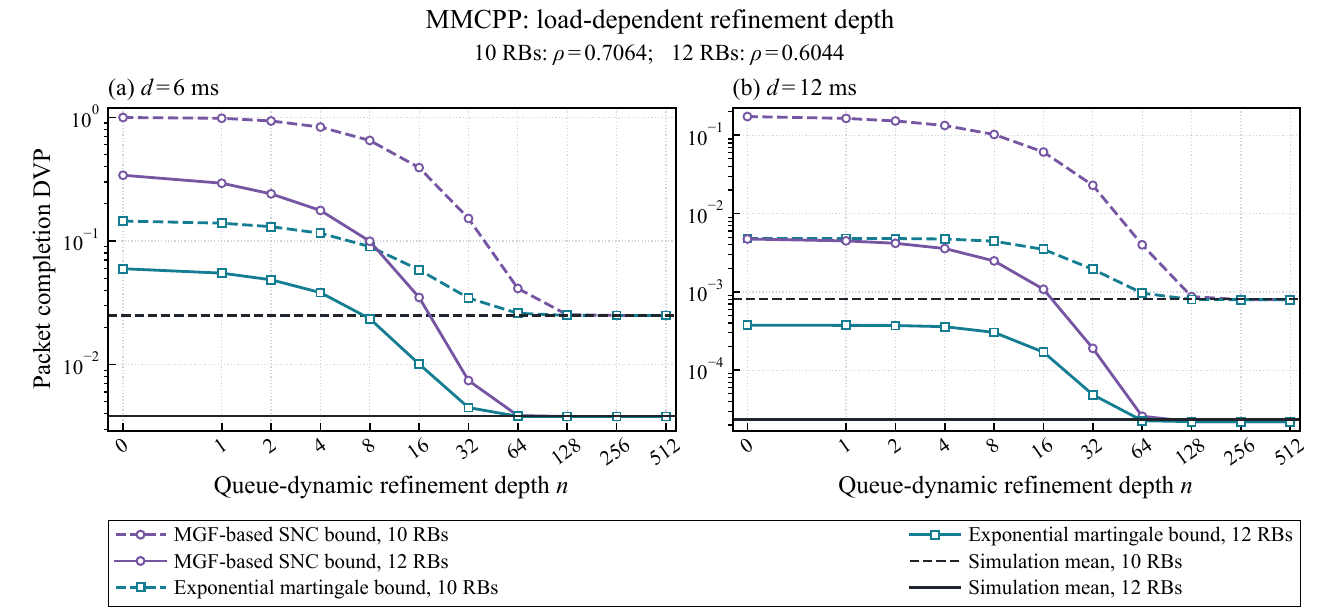}
\caption{MMCPP packet-DVP refinement at $10$ and $12$ RBs through $n=512$,
with (a) $d=6$ ms and (b) $d=12$ ms.}
\label{fig:markov_load_convergence}
\end{figure*}

\paragraph{Packet-delay hierarchies}
Figure~\ref{fig:markov_hierarchy} displays selected hierarchy levels
through $n=128$, approaching the empirical reference and extending the
scalar validation to dependent multistate dynamics.
Figure~\ref{fig:markov_load_convergence} extends the depth comparison
to $n=512$. The $12$-RB curves approach their empirical references around
$n=64$, compared with about $n=128$ at $10$ RBs. This depth difference also
appears in the GI/GI/1 baseline.

\subsection{AR($q$)-Driven Packetized Arrival with MM Wireless Service}
\label{subsec:arq_packetized_validation}

We take the state-dependent exponential-martingale analysis of
Gaussian AR($q$) traffic in~\cite{POLOCZEK201456} as our
methodological starting point. Zhu and Wang~\cite{zhu2024qos}
apply this type of construction to AR traffic with independent
block-fading service and QoS-aware radio resource allocation.
However, Gaussian AR($q$) values can be negative. Using them directly
as arrival increments would allow cumulative arrivals to decrease,
which cannot occur in a packet stream. We therefore use the Gaussian
process as a latent driver and construct a nonnegative budget before
packetization. We use blocks of length $H=1$ ms and define
\begin{equation}
\begin{aligned}
 \xi_k&=\sum_{\ell=1}^q\phi_\ell \xi_{k-\ell}+\varepsilon_k,
 \qquad \varepsilon_k\overset{\rm i.i.d.}{\sim}
                  \mathcal N(0,\sigma_\varepsilon^2),\\
 V_k&=b+\gamma \xi_k^2.
\end{aligned}
\label{eq:arq_packetized_budget}
\end{equation}
We use the stationary Gaussian source with AR coefficient vector
$\phi=(\phi_1,\ldots,\phi_q)$, whose entries are nonnegative and sum to less
than one, and set $\sigma_\varepsilon^2=1$.

To choose $b$ and $\gamma$, we start with a target mean budget $\mu_V$
in bits per block, set to match the common mean arrival rate. We assign
a fraction $1-\eta$ of this mean to the fixed budget $b=(1-\eta)\mu_V$
and the remaining fraction $\eta$ to the random budget $\gamma \xi_k^2$.
With $\eta=0.1$, the fixed and random parts account for $90\%$ and
$10\%$ of the mean budget, respectively. The stationary source has mean
zero, so $\mathbb E[\xi_k^2]=\operatorname{Var}(\xi_k)=s_\xi^2$.
We thus set $\gamma=\eta\mu_V/s_\xi^2$ to give the random part mean
$\eta\mu_V$ and the total budget mean $\mathbb E[V_k]=\mu_V$.
Since $\xi_k/s_\xi$ is standard normal, its square has a chi-square
distribution with one degree of freedom. Hence
$V_k\overset d=(1-\eta)\mu_V+\eta\mu_V\chi_1^2$, where $\chi_1^2$
denotes a random variable with this distribution.

A stationary renewal packetizer converts the accumulated budget
into complete packets~\cite{baccelli2003elements}. For each packet $i$,
we draw its required bit budget $L_i$ in advance from the packet-length
distribution specified in Subsection~\ref{subsec:common_wireless_setup}.
Within block $k$, budget accrues at rate $r(t)=V_k/H$; successive release
epochs satisfy
$T_i=\inf\{t>T_{i-1}:\int_{T_{i-1}}^t r(u)\,du\geq L_i\}$.
Any budget accumulated toward the next packet carries over across block
boundaries.

To examine how the strength of temporal dependence affects packet DVP
and refinement depth, we use the three AR(2) coefficient sets in
Table~\ref{tab:arq_dependence_profiles}. For each set, we recompute
$s_\xi^2$ and adjust $\gamma$ to keep the mean arrival rate fixed.
The table measures dependence by the lag-one budget autocorrelation
$r_V(1)=\operatorname{Corr}(\xi_k,\xi_{k-1})^2$.

\begin{table}[!htbp]
\centering
\caption{AR(2) Dependence Settings with Identical Marginal Budgets}
\label{tab:arq_dependence_profiles}
\begin{tabular}{lccc}
\toprule
Dependence & $\phi_1$ & $\phi_2$ & $r_V(1)$\\
\midrule
Weaker dependence   & 0.27 & 0.15 & 0.1009\\
Middle dependence   & 0.45 & 0.25 & 0.3600\\
Stronger dependence & 0.54 & 0.30 & 0.5951\\
\bottomrule
\end{tabular}
\end{table}

\paragraph{Conditional exponential-supermartingale bound}
At the reference block boundary, the predictive state
$Z_0=(\boldsymbol z,j,a)$ consists of the AR history
$\boldsymbol z=(\xi_{-q},\ldots,\xi_{-1})^{\mathsf T}$, the service state $j$,
and the budget $a$ already accumulated toward the next packet.
Let $L_{\max}$ be the maximum packet length in bits. We fix a finite
nonempty set $\Theta$ of admissible positive exponents before completion
and refinement. For each $\theta\in\Theta$, let $g_\theta(\boldsymbol z)$
and $c_\theta(j)$ be the source and service factors, respectively.
The conditional workload bound is
\begin{equation}
\begin{aligned}
 &\mathbb P(W_0>x\mid Z_0=(\boldsymbol z,j,a))
 \leq F(x,\boldsymbol z,j,a),\\
 &F(x,\boldsymbol z,j,a):=
 \min\!\left\{1,\min_{\theta\in\Theta}
 c_\theta(j)g_\theta(\boldsymbol z)
 e^{-\theta(x+a-L_{\max})}\right\}.
\end{aligned}
\label{eq:arq_conditional_workload_bound}
\end{equation}
Appendix~\ref{app:arq_conditional_bound} defines admissibility,
constructs the two factors, and proves the bound.

Even the exact conditional workload tail fails to decay uniformly across
predictive states, because a large budget in the preceding block can
leave substantial backlog at the reference boundary. Hence the uniform
tail-decay condition \eqref{eq:proper_initial_supersolution} in
Corollary~\ref{cor:remote_threshold_escape_convergence} cannot hold for
any valid workload envelope. Appendix~\ref{app:arq_conditional_bound}
uses the finite Gaussian second moment to verify the level-integrability
condition \eqref{eq:input_level_integrability} in
Proposition~\ref{prop:completion_nondegeneracy} at every level, so the
second alternative of the corollary applies to $U_0=\mathcal C(F)$.

A stationary budget origin lies uniformly in a length-biased renewal
gap. Given $a$, the pending packet length has law $L\mid L>a$ and,
together with the subsequent lengths, is independent of completed gaps.
The AR history $\boldsymbol z$ and service state $j$ determine the
conditional future source and service laws. Thus, given $Z_0$ under
$\mathbb P$, the block packet count and tagged contexts are independent
of $W_0$, verifying \eqref{eq:palm_state_compatibility} and
\eqref{eq:palm_boundary_separation} by \eqref{eq:packet_palm_identity}.

Let $f_\xi$ denote the $\mathcal N(0,s_\xi^2)$ density of $\xi_0$ under $\mathbb P$.
For the active source value $\xi_0=x$, averaging over the independent
stationary packetizer gives mean packet count
$(b+\gamma x^2)/\mathbb E[L]$ per block. The packet-count weighting
underlying \eqref{eq:packet_palm_boundary_state_law} yields the marginal
source density in the block containing the tagged release:
\begin{equation}
 f_\xi^{\rm pkt}(x)
 =\frac{b+\gamma x^2}{\mu_V}f_\xi(x)
 =\left[(1-\eta)+\eta\frac{x^2}{s_\xi^2}\right]f_\xi(x).
\label{eq:app_arq_packet_palm_source_law}
\end{equation}
Given $\xi_0$, the Gaussian history retains its ordinary conditional
law, and the independent service path retains its ordinary law.
Draw the tagged budget position $u^\star$ uniformly on $(0,V_0]$,
independently of the i.i.d.\ backward renewal gaps with the original
packet-length law, and set $\tau=Hu^\star/V_0$.
The first gap gives the tagged length, and tracing back to the first
renewal at or before zero jointly determines the earlier in-block
releases and the boundary age $a$.

\begin{figure*}[!htbp]
\centering
\includegraphics[width=\textwidth]{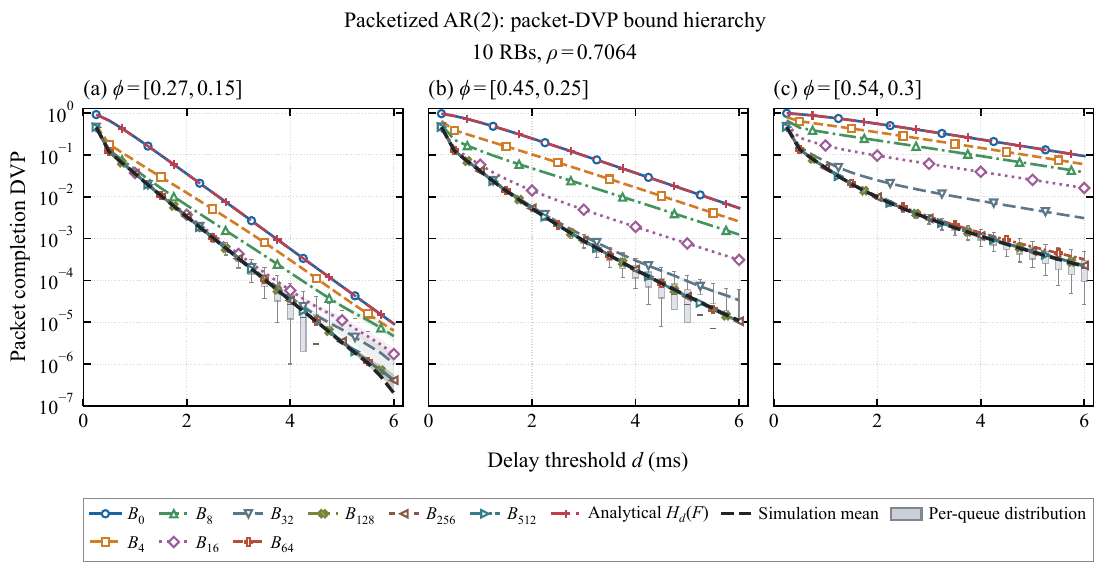}
\caption{AR(2) certified packet-DVP hierarchies at $10$ RBs through
$n=512$ for (a) weaker, (b) middle, and (c) stronger dependence, as
specified in Table~\ref{tab:arq_dependence_profiles}.}
\label{fig:arq_certified_hierarchy}
\end{figure*}

\paragraph{Dependence and refinement depth}
Figure~\ref{fig:arq_certified_hierarchy} shows the packet DVP bounds
for the three dependence settings at $10$ RBs through $n=512$.
Refinement tightens the bounds in all three settings.

\begin{figure*}[!htbp]
\centering
\includegraphics[width=\textwidth]{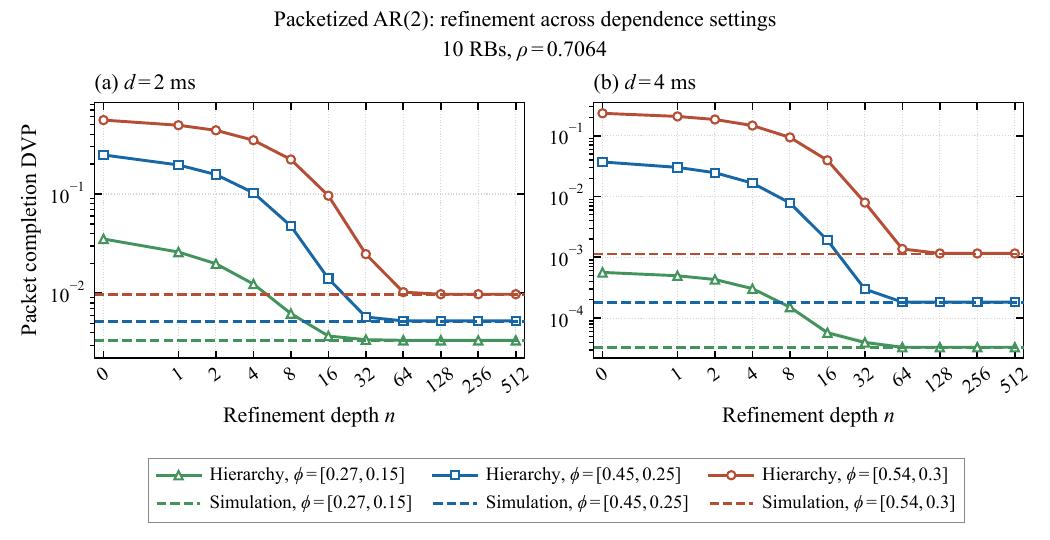}
\caption{AR(2) packet-DVP refinement across dependence settings at
$10$ RBs through $n=512$, with (a) $d=2$ ms and (b) $d=4$ ms.}
\label{fig:arq_coefficients_10rb}
\end{figure*}

Figure~\ref{fig:arq_coefficients_10rb} compares refinement at $d=2$
and $4$ ms. The bounds under weaker dependence approach their
empirical reference in fewer steps than those under stronger dependence.
At $n=512$, the refined bounds are close to their respective empirical
references for all three settings at both displayed thresholds.
These results illustrate refinement with a continuous predictive state.

\subsection{Poisson--Pareto Batch Arrivals with MM Wireless Service}
\label{subsec:poisson_pareto_batch_validation}

We construct heavy-tailed packet traffic using a compound-Poisson
process with discrete Pareto batch sizes. Its cumulative arrived
workload is $A(t)=\sum_{b=1}^{K(t)}B_b$ for $t\geq0$, with
$B_b=L_{b,\leq N_b}$,
where $K(t)$ is a Poisson counting process of rate $\lambda_b$,
$N_b$ is the number of packets in batch $b$, and the packet lengths
$L_{b,\ell}$ follow the common i.i.d. distribution specified above.
The i.i.d. batch sizes are independent of the batch epochs and packet
lengths, with $\mathbb P(N_b\geq n)=n^{-\alpha}$ for $n=1,2,\ldots$ and
$\alpha=1.5$. The tail-sum identity gives
$\mathbb E[N_b]=\sum_{n=1}^{\infty}\mathbb P(N_b\geq n)
=\sum_{n=1}^{\infty}n^{-\alpha}=\zeta(\alpha)$,
where $\zeta$ is the Riemann zeta function. We therefore choose
$\lambda_b=4000/\zeta(\alpha)$ batches/s to match the common mean
arrival rate of $2$ packets per TTI. The untruncated batch workload $B_b$ has
finite mean, infinite variance, and no finite positive exponential MGF.
The MGF-based SNC and exponential martingale workload bounds used above
require a finite arrival MGF at some positive exponent, so we instead
derive the following state-conditioned workload bound for heavy-tailed
bursts.

\paragraph{State-conditioned workload bound}
We use a constant reference service rate $c$ between the mean arrival
and mean service rates. The queue workload is bounded by the sum of
the backlog from the same arrivals under constant service at rate $c$
and the maximum cumulative shortfall of wireless service relative to
that rate.

Motivated by the heavy-tailed and self-similar (HTSS)
framework~\cite{liebeherr2012delay}, we bound the backlog tail with a
Pareto variable $X_c$ of shape $p=\alpha-1$ and scale $C_G(c)^{1/p}$,
where $C_G(c)$ is given by \eqref{eq:ppb_generator_power_bound}.
For each admissible pair $(c,\theta)$, we bound the service shortfall
tail with a shifted exponential variable $Z_{c,\theta,j}$ of rate
$\theta$. Its shift depends on the current service state $j$, and it
is chosen independently of $X_c$. The resulting conditional workload
bound is
\begin{equation}
\begin{aligned}
 \mathbb P(W_0>x\mid Z_0=j)&\leq F(x,j),\\
 F(x,j)&:=\min_{(c,\theta)\in\mathcal C}
 \mathbb P(X_c+Z_{c,\theta,j}>x).
\end{aligned}
\label{eq:ppb_convolution_workload_bound}
\end{equation}
Appendix~\ref{app:ppb_conditional_bound} gives the shift, defines
$\mathcal C$, and proves the bound.

Under the packet-Palm law in \eqref{eq:packet_palm_identity}, each
batch of size at least $k$ contributes one packet at rank $k$, so
$\mathbb P^A(r=k)=\mathbb P(N_b\geq k)/\mathbb E[N_b]
=k^{-\alpha}/\zeta(\alpha)$.
The future marked Poisson input is independent of the past and of
service, and the future service is independent of the past given the
full service state $Z_0=j$. This gives the conditional independence
required for \eqref{eq:palm_state_compatibility}
and \eqref{eq:palm_boundary_separation}.
Since $m(j)=H\lambda_b\mathbb E[N_b]$ is constant in $j$,
\eqref{eq:packet_palm_boundary_state_law} gives $\nu^A=\pi$.
The unconditional packet-DVP curves therefore follow from
\eqref{eq:packet_palm_state_disintegration} using the batch prefix
in \eqref{eq:palm_transfer_functional}.

\begin{figure*}[!htbp]
\centering
\includegraphics[width=\textwidth]{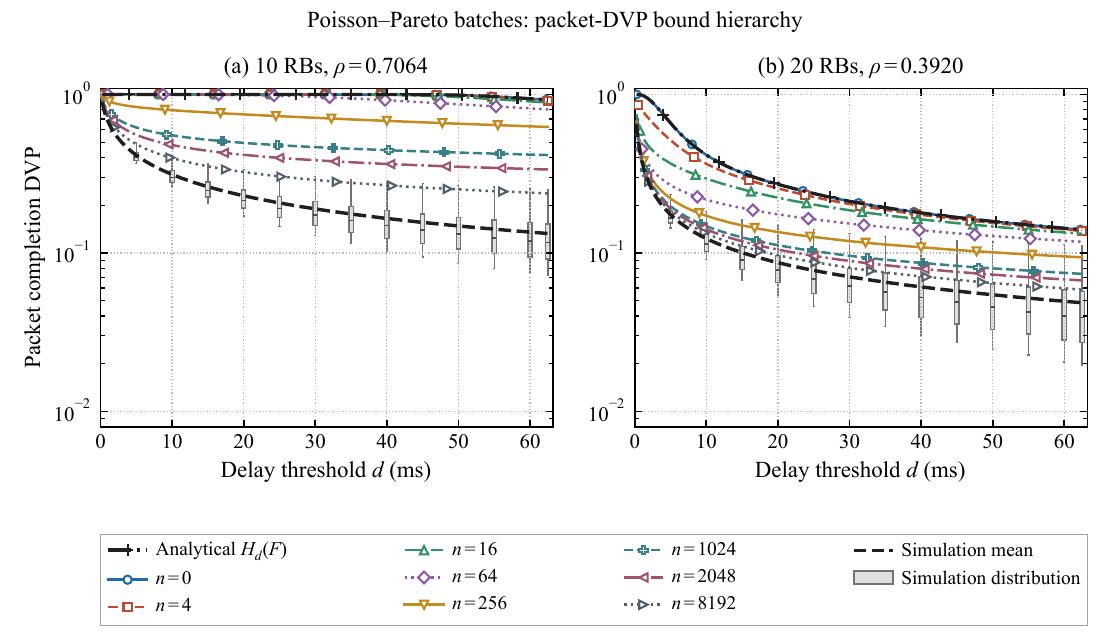}
\caption{Poisson--Pareto certified packet-DVP hierarchies through $n=8192$
at (a) $10$ RBs and (b) $20$ RBs.}
\label{fig:ppb_certified_hierarchy}
\end{figure*}

\begin{figure*}[!htbp]
\centering
\includegraphics[width=\textwidth]{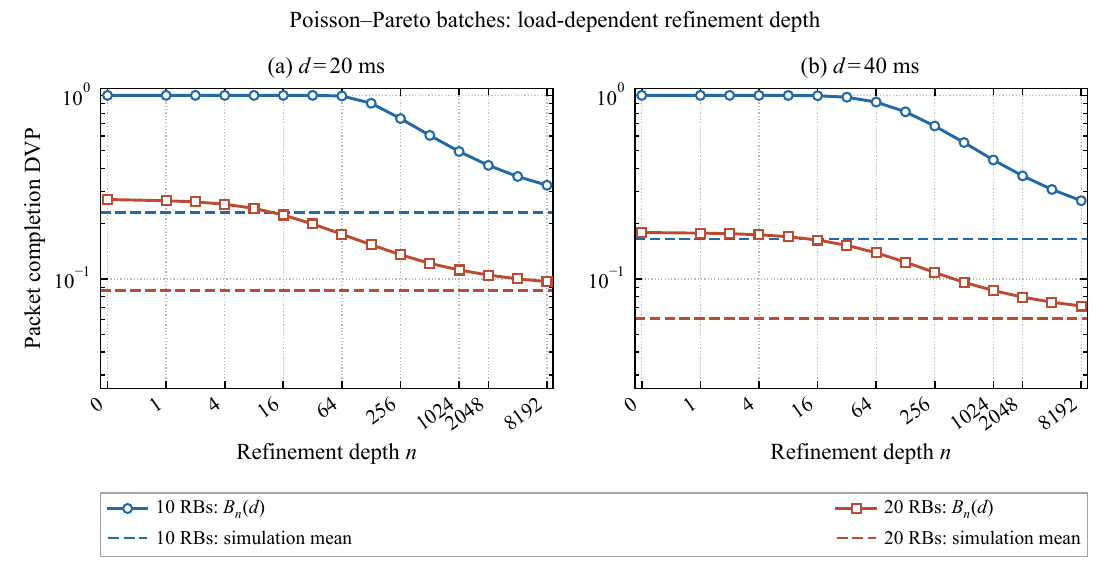}
\caption{Poisson--Pareto packet-DVP refinement at $10$ and $20$ RBs through
$n=8192$, with (a) $d=20$ ms and (b) $d=40$ ms.}
\label{fig:ppb_load_convergence}
\end{figure*}

\paragraph{Deep refinement under heavy-tailed bursts}
Figures~\ref{fig:ppb_certified_hierarchy}
and~\ref{fig:ppb_load_convergence} show that the Poisson--Pareto queue
requires much deeper refinement to approach its simulation references
than the GI/GI/1, MMCPP, and AR($q$) queues, whose curves are close to
their empirical references at $n=256$ in the preceding depth comparisons.
At both $10$ and $20$ RBs, the Poisson--Pareto bounds continue to tighten
from $n=2048$ to $8192$, with a gap to the simulation mean still visible
at $d=20$ and $40$ ms.
Despite this slower tightening, increasing the refinement depth
substantially reduces the conservatism of the starting bounds at both
allocations.

The error identity \eqref{eq:refinement_error_pathwise} evaluates the
initial gap $U_0-Q$ at the backward threshold $\tilde x_n$ along
continuing paths. Heavy-tailed bursts can absorb service surplus
accumulated over several blocks, keeping some continuing paths at
smaller thresholds for longer. This mechanism helps explain the slower tightening in
Fig.~\ref{fig:ppb_load_convergence}. At $20$ RBs, the larger mean service
surplus yields a larger asymptotic growth rate for $\tilde x_n$, consistent
with the tighter packet-DVP bounds relative to the simulation means.

\section{Conclusion}
\label{sec:conclusion}

This paper develops a framework for turning state-conditioned workload
bounds into certified hierarchies of decreasing conservatism. An exact
reflected block recursion yields a positive affine tail operator that
separates violations caused within a block from those that depend on
earlier workload. Least-supersolution completion, with a linear program
for exact finite-dimensional realizations, ensures that interpolating
adjacent levels gives continuously selectable supersolutions that can be
tightened further by the original operator. Under the stated conditions,
the hierarchy converges to the stationary workload tail. Exact packet-Palm
transfer carries the hierarchy and its error propagation to
packet-completion delay. Numerical results illustrate how refinement depth
controls conservatism across wireless queueing models.

Extending this single-hop framework to tandem networks requires a joint
state that captures how upstream departures drive downstream arrivals,
together with a family of threshold events that remains closed under the
queue dynamics. Cumulative workloads across successive hops offer a
starting point for fluid tandems. For store-and-forward packet networks,
the main challenges are to retain packet order without making the joint
state intractable, and to establish convergence and an exact transfer to
end-to-end packet delay.

\clearpage
\appendices
\section{Conditional Bounds for the MMCPP Queue}
\label{app:markov_conditional_bounds}

\subsection{Backward representation and MGF bound}
Starting from the current boundary state $J_0=z$, we enumerate preceding
blocks from the most recent to the oldest by $m=1,2,\ldots$.
The backward states $J_m$ evolve with $\check P$. Write $\Pr_z$ and
$\mathbb E_z$ for probability and expectation conditional on $J_0=z$,
and $\mathbb E_\pi$ for expectation with $J_0\sim\pi$.
Let $A_m$ and $S_m$ denote the total arrived workload and available
service in historical block $m$. Conditional on the state sequence,
distinct block marks are independent. Within block $m$, $A_m$ and $S_m$
are independent with transforms $a_{J_m}(\theta)$ and $s_{J_m}(\theta)$.
Write $Y_m=A_m-S_m$ and define
$\widetilde S_0=0$ and $\widetilde S_m=\sum_{\ell=1}^mY_\ell$.
Unrolling \eqref{eq:reflected_block_recursion} over the past and taking
the stable stationary backward limit gives
\begin{equation}
\begin{aligned}
 W_0
 &=\sup_{m\geq1}\{\widetilde S_{m-1}+R_m\}\\
 &=\sup\{R_1,\,Y_1+R_2,\,Y_1+Y_2+R_3,\ldots\}.
\end{aligned}
\label{eq:app_markov_backward_workload}
\end{equation}
Here $R_m$ is generated at the right boundary of the $m$th historical
block, and $\widetilde S_{m-1}$ accumulates the net increments of the
$m-1$ more recent blocks. Since $R_m\leq A_m$,
\begin{equation}
 \{W_0>x\}\subseteq
 \bigcup_{m\geq1}\{\widetilde S_{m-1}+A_m>x\}.
\label{eq:app_markov_restart_union}
\end{equation}
For each term, the conditional Chernoff inequality gives, for $\theta>0$,
\begin{equation}
\begin{aligned}
 &\Pr_z(\widetilde S_{m-1}+A_m>x)\\
 &\qquad\leq e^{-\theta x}
       \mathbb E_z[e^{\theta(\widetilde S_{m-1}+A_m)}].
\end{aligned}
\label{eq:app_markov_conditional_chernoff}
\end{equation}

To evaluate this expectation, define the lifted transform vectors
\[
 a_\theta=\bigl(a_j(\theta)\bigr)_{j\in\mathsf Z},
 \qquad
 s_\theta=\bigl(s_j(\theta)\bigr)_{j\in\mathsf Z}.
\]
Then set
\begin{equation}
 b_\theta=\check P a_\theta,\qquad
 G_\theta=\check P\operatorname{diag}(a_\theta s_\theta),
\label{eq:app_markov_mgf_kernel}
\end{equation}
where the product $a_\theta s_\theta$ is componentwise.
For the first historical block,
$\mathbb E_z[e^{\theta A_1}]
=\sum_j\check P_{zj}a_j(\theta)=[b_\theta]_z$.
For two blocks, conditioning on $J_1=j$ yields
\begin{equation}
\begin{aligned}
 \mathbb E_z[e^{\theta(Y_1+A_2)}]
 &=\sum_j\check P_{zj}a_j(\theta)s_j(\theta)[b_\theta]_j\\
 &=[G_\theta b_\theta]_z.
\end{aligned}
\label{eq:app_markov_two_block_mgf}
\end{equation}
Repeating this conditioning step proves
\begin{equation}
 \mathbb E_z[e^{\theta(\widetilde S_{m-1}+A_m)}]
 =[G_\theta^{m-1}b_\theta]_z.
\label{eq:app_markov_history_mgf}
\end{equation}
Thus each of the $m-1$ more recent blocks contributes its net-input
transform, whereas the oldest block contributes only its arrival
transform. The backward transition precedes the destination block mark,
which fixes the matrix order in \eqref{eq:app_markov_mgf_kernel}.

Applying the union bound to the pathwise inclusion
in \eqref{eq:app_markov_restart_union}, and then using
\eqref{eq:app_markov_conditional_chernoff} and
\eqref{eq:app_markov_history_mgf}, gives
\begin{equation}
\begin{aligned}
 Q(x,z)
 &\leq \sum_{m\geq1}
       \Pr_z(\widetilde S_{m-1}+A_m>x)\\
 &\leq e^{-\theta x}
       \sum_{m\geq1}
       \mathbb E_z[e^{\theta(\widetilde S_{m-1}+A_m)}]\\
 &=e^{-\theta x}
       \sum_{m\geq1}[G_\theta^{m-1}b_\theta]_z.
\end{aligned}
\label{eq:app_markov_infinite_history_bound}
\end{equation}
It remains to sum the contributions from infinitely many historical
blocks. Whenever $\operatorname{spr}(G_\theta)<1$, the matrix geometric
series converges, and the Neumann series identity
\cite{horn2013matrix} yields
\begin{equation}
 Q(x,z)\leq
 e^{-\theta x}[(I-G_\theta)^{-1}b_\theta]_z.
\label{eq:app_markov_resolvent_derivation}
\end{equation}

Positive exponents satisfying this convergence condition exist
under the present stable, light-tailed model.
Indeed, let $r(\theta)=\operatorname{spr}(G_\theta)$.
The finite irreducible state chain and block MGFs finite in a neighborhood
of zero ensure differentiability of the Perron root there.
At $G_0=\check P$, the left and right Perron eigenvectors are
$\pi$ and $\mathbf1$, respectively, with $\pi^{\mathsf T}\mathbf1=1$.
The simple-eigenvalue derivative formula
\cite{greenbaum2020firstorder} therefore gives
\begin{equation}
\begin{aligned}
 r(0)&=1,\\
 r'(0)&=\pi^{\mathsf T}G'_0\mathbf1
       =\mathbb E_\pi[A_m-S_m]<0.
\end{aligned}
\label{eq:app_markov_spectral_drift}
\end{equation}
Thus $r(\theta)<1$ for all sufficiently small positive $\theta$.
Moreover, the nonzero entries of $G_\theta$ are log-convex,
so the spectral-radius log-convexity theorem
\cite{nussbaum1986convexity}
implies that $\log r(\theta)$ is convex.
Consequently, the admissible set
$\{\theta>0:r(\theta)<1\}$ is a nonempty interval.

Taking the infimum of \eqref{eq:app_markov_resolvent_derivation}
over this interval and capping by one proves
\eqref{eq:markov_classical_bound}.
The derivation follows the union-bound and Chernoff approach
of classical MGF-based SNC~\cite{bouillard2024mgf},
while retaining conditioning on the current predictive state
and explicitly accounting for the within-block restart term.

\subsection{Conditional martingale bound}

We adapt the exponential-martingale stopping argument
of~\cite{10.1145/3287322} to the upper comparison
\begin{equation}
 W_0\leq\sup_{m\geq1}V_m,
 \qquad V_m:=\widetilde S_{m-1}+A_m,
\label{eq:app_markov_martingale_comparison}
\end{equation}
which follows from \eqref{eq:app_markov_backward_workload}
and $R_m\leq A_m$.
Let $h_\theta>0$ be a right Perron eigenvector of $G_\theta$,
and define
$H_\theta=\min_{j\in\mathsf Z}
\{s_j(\theta)h_\theta(j)\}>0$.
Consider the process
\begin{equation}
\begin{aligned}
 M_0&=h_\theta(J_0),\\
 M_m&=e^{\theta V_m}s_{J_m}(\theta)h_\theta(J_m),
 \qquad m\geq1,
\end{aligned}
\label{eq:app_markov_single_index_martingale}
\end{equation}
with $\mathcal F_0=\sigma(J_0)$ and
\begin{equation}
 \mathcal F_m=
 \sigma\!\left(
 J_0,(J_\ell,A_\ell)_{\ell=1}^{m},
 (S_\ell)_{\ell=1}^{m-1}
 \right),
 \qquad m\geq1.
\label{eq:app_markov_single_index_filtration}
\end{equation}
Thus $V_m$ and $J_m$ are known under $\mathcal F_m$,
while $S_m$ is not included.

By the admissibility analysis in the preceding subsection,
$r(\theta)=\operatorname{spr}(G_\theta)\leq1$
for $0<\theta\leq\theta^\star$.
The Perron relation therefore gives
$G_\theta h_\theta=r(\theta)h_\theta\leq h_\theta$.
Using $V_{m+1}=V_m-S_m+A_{m+1}$, conditional independence
and the backward transition kernel yield
\begin{equation}
\begin{aligned}
 &\mathbb E_z[M_{m+1}\mid\mathcal F_m]\\
 &\quad=e^{\theta V_m}s_{J_m}(\theta)
          (G_\theta h_\theta)(J_m)\\
 &\quad=r(\theta)M_m\leq M_m.
\end{aligned}
\label{eq:app_markov_single_index_condition}
\end{equation}
The initial step likewise satisfies
$\mathbb E_z[M_1\mid\mathcal F_0]
=r(\theta)M_0\leq M_0$.
Hence $M$ is a nonnegative supermartingale throughout
this parameter range.

Let $\tau_x=\inf\{m\geq1:V_m>x\}$, with
$\inf\varnothing=\infty$.
This is a stopping time for $(\mathcal F_m)$, and
\eqref{eq:app_markov_martingale_comparison} implies
$Q(x,z)\leq\Pr_z(\tau_x<\infty)$.
On $\{\tau_x<\infty\}$,
$M_{\tau_x}>H_\theta e^{\theta x}$.
Applying optional stopping~\cite{williams1991probability} at $\tau_x\wedge N$
for a finite integer $N\geq1$ gives
\begin{equation}
\begin{aligned}
 h_\theta(z)
 &\geq\mathbb E_z[M_{\tau_x\wedge N}]\\
 &\geq\mathbb E_z[
       M_{\tau_x}\mathbf1_{\{\tau_x\leq N\}}]\\
 &\geq H_\theta e^{\theta x}\Pr_z(\tau_x\leq N).
\end{aligned}
\label{eq:app_markov_stopped_inequality}
\end{equation}
Letting $N\to\infty$ and capping by one yields
\begin{equation}
\begin{gathered}
 Q(x,z)\leq
 \min\!\left\{
 1,\frac{h_\theta(z)}{H_\theta}e^{-\theta x}
 \right\},
 \\
 0<\theta\leq\theta^\star.
\end{gathered}
\label{eq:app_markov_conditional_stopping_bound}
\end{equation}
Taking $\theta=\theta^\star$ gives
\eqref{eq:markov_martingale_bound}.
The minimum defining $H_\theta$ is taken over all predictive
states because arrival and service amounts remain random
conditional on the modulating state, so the state alone
does not determine whether an upcrossing occurs.

\section{Conditional Bound for Squared-Gaussian AR($q$) Packetized Traffic}
\label{app:arq_conditional_bound}

\subsection{Exact conditional Gaussian transform}
For $m\geq0$, write
\[
 \boldsymbol z_m=(\xi_{-m-q},\ldots,\xi_{-m-1})^{\mathsf T},
 \qquad \boldsymbol z_0=\boldsymbol z.
\]
A stationary scalar Gaussian process has the same law under time
reversal~\cite{weiss1975reversibility}. Hence
\[
 \boldsymbol z_{m+1}=B\boldsymbol z_m+e_1\widetilde\varepsilon_{m+1},
 \qquad \widetilde\varepsilon_{m+1}\sim
          \mathcal N(0,\sigma_\varepsilon^2),
\]
where $B$ has first row $(\phi_1,\ldots,\phi_q)$ and unit
subdiagonal, and $e_1$ is the first coordinate vector. The backward
innovations are independent of the already revealed source history.
Write $\boldsymbol z'$ for the next backward source state given
the current state $\boldsymbol z$. The next historical budget is
$v(\boldsymbol z_m)=b+\gamma(z_m)_q^2$.

Let $R_\theta\in\mathbb R^{q\times q}$ be a real symmetric
positive semidefinite matrix. Define
$g_\theta(\boldsymbol z)=e^{\boldsymbol z^{\mathsf T}R_\theta\boldsymbol z}$
and require $d_\theta=1-2\sigma_\varepsilon^2e_1^{\mathsf T}R_\theta e_1>0$.
Completing the square in the scalar Gaussian innovation gives
\begin{equation}
\begin{aligned}
 &\mathbb E[g_\theta(\boldsymbol z')\mid\boldsymbol z]\\
 &\quad=d_\theta^{-1/2}
 \exp\!\left\{\boldsymbol z^{\mathsf T}
 \left(B^{\mathsf T}R_\theta B+
 \frac{2\sigma_\varepsilon^2}{d_\theta}
 B^{\mathsf T}R_\theta e_1e_1^{\mathsf T}R_\theta B\right)
 \boldsymbol z\right\}.
\end{aligned}
\label{eq:app_arq_gaussian_transform}
\end{equation}
With $e_q$ the last coordinate vector, define
\begin{equation}
\begin{aligned}
 D_\theta={}&R_\theta-\theta\gamma e_qe_q^{\mathsf T}
             -B^{\mathsf T}R_\theta B\\
 &-\frac{2\sigma_\varepsilon^2}{d_\theta}
        B^{\mathsf T}R_\theta e_1e_1^{\mathsf T}R_\theta B.
\end{aligned}
\label{eq:app_arq_certificate_parameters}
\end{equation}
If $D_\theta\succeq0$, the exact identity
\begin{equation}
\begin{aligned}
 e^{\theta v(\boldsymbol z)}
 \mathbb E[g_\theta(\boldsymbol z')\mid\boldsymbol z]
 &=\alpha_\theta g_\theta(\boldsymbol z)
        e^{-\boldsymbol z^{\mathsf T}D_\theta\boldsymbol z}\\
 &\leq\alpha_\theta g_\theta(\boldsymbol z),\qquad
 \alpha_\theta=\frac{e^{\theta b}}{\sqrt{d_\theta}}
\end{aligned}
\label{eq:app_arq_exact_arrival_certificate}
\end{equation}
holds for every $\boldsymbol z\in\mathbb R^q$.
For fixed $\theta$, $D_\theta=0$ is a discrete algebraic Riccati
equation~\cite{lancaster1995algebraic}. This equality, or its
matrix-inequality version $D_\theta\succeq0$, certifies the source
transform globally.

\subsection{Service coefficients and the stopping bound}
Let $J_m$ be the backward six-state service chain, with $J_0=j$
and transition matrix $\check P_S$, the stationary reversal of $P_S$.
Let $S_m$ denote the service in historical block $m$. Define
\[
\begin{aligned}
 s_j(\theta)&:=\mathbb E[e^{-\theta S_m}\mid J_m=j],\\
 K_S(\theta)&:=\check P_S\operatorname{diag}(s_j(\theta)).
\end{aligned}
\]
Here $K_S(\theta)$ is the nonnegative service transform matrix,
with entries $[K_S(\theta)]_{ij}=\check P_S(i,j)s_j(\theta)$.

For a positive exponent satisfying the arrival certificate,
we seek a finite vector $c_\theta$ satisfying
\begin{equation}
 c_\theta\geq\mathbf1,\qquad
 \alpha_\theta K_S(\theta)c_\theta\leq c_\theta.
\label{eq:app_arq_service_supervector}
\end{equation}
The arrival factor scales the service kernel, so
$\operatorname{spr}(\alpha_\theta K_S(\theta))
=\alpha_\theta\lambda_S(\theta)$, where
$\lambda_S(\theta)=\operatorname{spr}(K_S(\theta))$.
If $\alpha_\theta\lambda_S(\theta)\leq1$, a positive right Perron
vector of $K_S(\theta)$ normalized by its minimum is feasible.
We reduce the service prefactor by solving
\begin{equation}
 \min_c\ \mathbf1^{\mathsf T}c
 \quad\text{s.t.}\quad
 c\geq\mathbf1,\quad\alpha_\theta K_S(\theta)c\leq c.
\label{eq:app_arq_service_coefficient_selection}
\end{equation}
The feasible set is closed under componentwise minima because
$K_S(\theta)$ is nonnegative. Thus this choice is no larger
componentwise than the normalized Perron choice.

Define the admissible exponent set for the arrival and service
certificates by
\begin{equation}
\begin{aligned}
 \mathcal D:=\{\theta>0:\;&R_\theta\succeq0,\quad d_\theta>0,\\
 &D_\theta\succeq0,\quad
   \alpha_\theta\lambda_S(\theta)\leq1\}.
\end{aligned}
\label{eq:app_arq_exponent_selection}
\end{equation}
Write $\bar s=\mathbb E_{\pi_S}[S_m]$. Since $B$ is stable,
there is a smooth local Riccati solution $R_\theta\succeq0$
through $R_0=0$, with $\alpha_0=\lambda_S(0)=d_0=1$.
Differentiating the Riccati equation and using source stationarity,
together with Perron differentiation of the service kernel, gives
\[
\begin{aligned}
 \log\alpha_\theta&=\mu_V\theta+o(\theta),\\
 \log\lambda_S(\theta)&=-\bar s\theta+o(\theta).
\end{aligned}
\]
Hence
\[
 \log[\alpha_\theta\lambda_S(\theta)]
 =(\mu_V-\bar s)\theta+o(\theta)<0
\]
for sufficiently small positive $\theta$ under mean stability
$\mu_V<\bar s$. The local Riccati certificate and continuity
of $d_\theta$ therefore imply that there exists $\epsilon>0$
such that $(0,\epsilon)\subseteq\mathcal D$.
For finite thresholds, we retain a fixed finite nonempty family
$\Theta\subset\mathcal D$ and take the minimum
in~\eqref{eq:arq_conditional_workload_bound}.

Set
\[
\begin{aligned}
 \widetilde S_0^{\mathrm f}&=0,\qquad
 \widetilde S_n^{\mathrm f}=\sum_{m=1}^n[v(\boldsymbol z_{m-1})-S_m],\\
 M_n&=e^{\theta \widetilde S_n^{\mathrm f}}g_\theta(\boldsymbol z_n)c_\theta(J_n).
\end{aligned}
\]
Let $\mathcal F_n$ be generated by $(\boldsymbol z_0,J_0)$ and
the triples $(\boldsymbol z_m,J_m,S_m)$, $1\leq m\leq n$.
Source--service
independence and~\eqref{eq:app_arq_exact_arrival_certificate} give
\begin{equation}
\begin{aligned}
 \mathbb E[M_{n+1}\mid\mathcal F_n]
 &=e^{\theta \widetilde S_n^{\mathrm f}}e^{\theta v(\boldsymbol z_n)}
   \mathbb E[g_\theta(\boldsymbol z_{n+1})\mid\boldsymbol z_n]
   [K_S(\theta)c_\theta]_{J_n}\\
 &\leq e^{\theta \widetilde S_n^{\mathrm f}}g_\theta(\boldsymbol z_n)
       [\alpha_\theta K_S(\theta)c_\theta]_{J_n}
 \leq M_n.
\end{aligned}
\label{eq:app_arq_joint_supermartingale}
\end{equation}
Hence $M_n$ is a nonnegative supermartingale with
$M_0=g_\theta(\boldsymbol z)c_\theta(j)$.

Consider the auxiliary fluid queue fed by the budget clock before
packetization. Both its input and the wireless service are linear
within each coherence block, so the backward reflected supremum
is attained at a block boundary:
$W_f(0)=\sup_{n\geq0}\widetilde S_n^{\mathrm f}$.
For $y\geq0$, define $\tau_y=\inf\{n\geq0:\widetilde S_n^{\mathrm f}>y\}$.
At an upcrossing, $g_\theta\geq1$ and $c_\theta\geq\mathbf1$
imply $M_{\tau_y}\geq e^{\theta y}$.
Bounded optional stopping~\cite{williams1991probability} gives
\[
 g_\theta(\boldsymbol z)c_\theta(j)
 \geq\mathbb E[M_{\tau_y\wedge N}\mid\boldsymbol z,j]
 \geq e^{\theta y}\Pr(\tau_y\leq N\mid\boldsymbol z,j).
\]
Letting $N\to\infty$ proves
\begin{equation}
 \Pr(W_f(0)>y\mid\boldsymbol z,j)
 \leq\min\{1,c_\theta(j)g_\theta(\boldsymbol z)e^{-\theta y}\}.
\label{eq:app_arq_fluid_stopping_bound}
\end{equation}

Let $a(t)$ denote the budget accumulated since the last packet
release. The independent stationary renewal construction in
budget coordinates makes $a=a(0)$ independent of the source and
channel histories, so~\eqref{eq:app_arq_fluid_stopping_bound}
remains valid conditional on $a$.

Write $A_{\rm pkt}(s,t)$ for the total workload of packets released in
$(s,t]$ and $A_f(s,t)$ for the budget accumulated over the same interval.
For $t\geq0$, packetizer conservation over $(0,t]$ gives
\begin{equation}
\begin{aligned}
 A_{\rm pkt}(0,t)
 &=A_f(0,t)+a(0)-a(t)\\
 &\leq A_f(0,t)+L_{\max}-a(t).
\end{aligned}
\label{eq:app_arq_packetizer_conservation}
\end{equation}
The same identity holds on every subinterval. Taking reflected
suprema with the same service and a common observation endpoint,
then placing that stationary endpoint at zero, gives the pathwise
comparison $W_0\leq W_f(0)+L_{\max}-a$.
For $y=x+a-L_{\max}\geq0$, applying
\eqref{eq:app_arq_fluid_stopping_bound} conditional on
$(\boldsymbol z,j,a)$ proves the corresponding member of
\eqref{eq:arq_conditional_workload_bound}. Taking the minimum over the fixed family
$\Theta$ gives the stated conditional bound.

The workload recursion \eqref{eq:reflected_block_recursion} and the
conservation equality in
\eqref{eq:app_arq_packetizer_conservation} over the most recent
historical block $(-H,0]$, with $a(-H)\geq0$ and $a(0)\leq L_{\max}$,
give
\[
 W_0\geq A_{\rm pkt}(-H,0)-S_1
     \geq b+\gamma z_q^2-L_{\max}-S_1,
\]
where $z_q=(\boldsymbol z)_q=\xi_{-1}$.
Conditional on $Z_0=(\boldsymbol z,j,a)$, the law of $S_1$ depends only
on $j$. Since $\gamma>0$ and $S_1$ is finite almost surely, for each
fixed finite $x\geq0$ the conditional probability
that the last expression exceeds $x$ tends to one as $|z_q|\to\infty$,
uniformly over the finitely many service states $j$. Thus, for any
$\delta\in(0,1)$, there is a finite $K$
such that $Q(x,\cdot)>1-\delta$ holds $\pi$-almost surely on
$\{|z_q|\geq K\}$. This event has positive $\pi$-probability because
$z_q$ has an unbounded Gaussian marginal. Consequently,
$\operatorname{ess\,sup}_{\pi}Q(x,\cdot)=1$ for every finite $x\geq0$,
which rules out \eqref{eq:proper_initial_supersolution} for any valid
workload envelope $U_0\geq Q$.

The level-integrability condition can be checked directly. Fix any
$\theta\in\Theta$. Since $a\geq0$,
\eqref{eq:arq_conditional_workload_bound} implies
\[
\begin{aligned}
 \ell_\varepsilon(\boldsymbol z,j,a)
 &\leq1+L_{\max}\\
 &\quad+\theta^{-1}\!\left[
 \log c_\theta(j)+\boldsymbol z^{\mathsf T}R_\theta\boldsymbol z
                  +\log(1/\varepsilon)\right].
\end{aligned}
\]
The service state space is finite, and the stationary Gaussian history
$\boldsymbol z$ has a finite covariance matrix
$\Sigma_{\boldsymbol z}:=\operatorname{Cov}(\boldsymbol z)$. Hence
$\mathbb E[\boldsymbol z^{\mathsf T}R_\theta\boldsymbol z]
=\operatorname{tr}(R_\theta\Sigma_{\boldsymbol z})<\infty$.
Thus \eqref{eq:input_level_integrability} holds at every
$\varepsilon\in(0,1)$, verifying the input condition of
Proposition~\ref{prop:completion_nondegeneracy} and
Corollary~\ref{cor:remote_threshold_escape_convergence}.

\section{Conditional Bound for Poisson--Pareto Batch Arrivals}
\label{app:ppb_conditional_bound}

Let $B$ and $N$ be the workload and packet count of a generic batch.
For blocks of duration $H=1$ ms, choose a reference rate
$r<c<\bar s$, where $r=\lambda_b\mathbb E[B]$ and
$\bar s=\mathbb E_{\pi_S}[S_m]/H$ are the mean arrival and service rates.
With the current boundary at time zero, split the past net input as
$A(-t,0)-S(-t,0)=[A(-t,0)-ct]+[ct-S(-t,0)]$ for $t\geq0$ and define
\[
\begin{aligned}
 W_A^{(c)}&:=\sup_{t\geq0}\{A(-t,0)-ct\},\\
 D_S^{(c)}&:=\sup_{t\geq0}\{ct-S(-t,0)\}.
\end{aligned}
\]
Here $(-t,0]$ denotes an actual past interval of length $t$;
these suprema retain the historical paths conditional on $J_0=j$.
Taking suprema gives the pathwise comparison
\begin{equation}
 W_0\leq W_A^{(c)}+D_S^{(c)}.
\label{eq:app_ppb_pathwise_split}
\end{equation}
Here $W_A^{(c)}$ is the workload of the batch-arrival queue served
at constant rate $c$, while $D_S^{(c)}$ is the maximum cumulative
service deficit relative to $ct$. Arrival--channel independence
makes these two random variables independent conditional on the
current predictive service state $J_0=j$, without changing the law
of $W_A^{(c)}$. We therefore bound the two terms separately and
combine their conditional distributions by convolution.

\subsection{Arrival-side workload bound}
Since $B\leq L_{\max}N$, for every $u>0$,
\[
\begin{aligned}
 \mathbb P(B>u)
 &\leq\mathbb P(N>u/L_{\max})\\
 &=\mathbb P(N\geq\lfloor u/L_{\max}\rfloor+1)\\
 &=(\lfloor u/L_{\max}\rfloor+1)^{-\alpha}\\
 &\leq c_Bu^{-\alpha},\qquad c_B=L_{\max}^{\alpha}.
\end{aligned}
\]
The second equality uses the model definition
$\mathbb P(N\geq n)=n^{-\alpha}$.
Write $\mu_B=\mathbb E[B]=\mathbb E[N]\mathbb E[L]<\infty$
and $p=\alpha-1$.
The equilibrium residual batch workload $B_e$ has tail
\begin{equation}
\begin{aligned}
 \mathbb P(B_e>x)
 &=\frac1{\mu_B}\int_x^\infty\mathbb P(B>u)\,du\\
 &\leq\frac{c_B}{p\mu_B}x^{-p},\qquad x>0.
\end{aligned}
\label{eq:app_ppb_residual_tail}
\end{equation}

For the auxiliary queue defined above, let $\rho_c=r/c\in(0,1)$.
Grouping each batch into one task with
service time $B/c$ preserves the total unfinished bits and gives a
stable M/G/1 queue. The exact Pollaczek--Khinchine representation
\cite{mandjes2023cramer} therefore gives
\begin{equation}
 W_A^{(c)}\overset d=\sum_{k=1}^{G_c}B_{e,k},
\label{eq:app_ppb_geometric_sum}
\end{equation}
where $\mathbb P(G_c=n)=(1-\rho_c)\rho_c^n$, $n\geq0$,
and the $B_{e,k}$ are i.i.d. copies of $B_e$, independent of $G_c$.
The empty sum is zero.

For any deterministic weights $w_k>0$ with
$\sum_{k\geq1}w_k=1$, the following pathwise inclusion holds for $x>0$:
\[
 \left\{\sum_{k=1}^{G_c}B_{e,k}>x\right\}
 \subseteq
 \bigcup_{k\geq1}\{G_c\geq k,\ B_{e,k}>w_kx\}.
\]
Indeed, if no term on the right occurs, the random sum is at most
$x\sum_{k=1}^{G_c}w_k\leq x$.
The union bound, independence, and
\eqref{eq:app_ppb_residual_tail} therefore give
\[
\begin{aligned}
 \mathbb P(W_A^{(c)}>x)
 &\leq\sum_{k\geq1}\rho_c^k\mathbb P(B_e>w_kx)\\
 &\leq\frac{c_Bx^{-p}}{p\mu_B}
             \sum_{k\geq1}\rho_c^kw_k^{-p}.
\end{aligned}
\]
To control the coefficient, H\"older's inequality~\cite{hardy1952inequalities}
with conjugate exponents $p+1$ and $(p+1)/p$ yields
\[
\begin{aligned}
 \sum_{k\geq1}\rho_c^{k/(p+1)}
 &=\sum_{k\geq1}(\rho_c^kw_k^{-p})^{1/(p+1)}
                         w_k^{p/(p+1)}\\
 &\leq\left(\sum_{k\geq1}\rho_c^kw_k^{-p}\right)^{1/(p+1)},
\end{aligned}
\]
where $\sum_kw_k=1$ was used. Equality holds when
$w_k\propto\rho_c^{k/(p+1)}$.
Since $p+1=\alpha$, the normalized geometric weights
$w_k=(1-\rho_c^{1/\alpha})\rho_c^{(k-1)/\alpha}$ attain the
smallest coefficient in this estimate and permit geometric summation.
Substitution gives
\begin{equation}
\begin{aligned}
 \mathbb P(W_A^{(c)}>x)
 &\leq\frac{c_B\rho_c}{p\mu_B(1-\rho_c^{1/\alpha})^\alpha}x^{-p}\\
 &=C_G(c)x^{-p}.
\end{aligned}
\label{eq:app_ppb_arrival_power_bound}
\end{equation}
Substituting $\rho_c=r/c$ and capping the bound at one gives
$\mathbb P(W_A^{(c)}>x)\leq g_c(x)$ for $x\geq0$, where
\begin{equation}
\begin{aligned}
 g_c(x)&=\min\{1,C_G(c)x^{-p}\},\quad x>0,\qquad g_c(0)=1,\\
 C_G(c)&=\frac{c_B}
 {p\mu_B[(c/r)^{1/\alpha}-1]^{\alpha}}.
\end{aligned}
\label{eq:ppb_generator_power_bound}
\end{equation}

\subsection{Conditional service deficit and convolution}
For the backward predictive service chain, including its clock phases, define
$s_j(\theta)=\mathbb E[e^{-\theta S_m}\mid J_m=j]$ and
$K_S(\theta)=\check P_S\operatorname{diag}(s_j(\theta))$,
with $\lambda_S(\theta)=\operatorname{spr}(K_S(\theta))$.
For the reference rate $c$ fixed above, choose $\theta>0$ satisfying
$e^{\theta cH}\lambda_S(\theta)\leq1$.
Existence follows from $c<\bar s$ by the Perron drift argument in
Appendix~\ref{app:markov_conditional_bounds}.
Let $v_\theta>0$ be a right Perron eigenvector normalized by
$\min_jv_\theta(j)=1$. Conditional on $J_0=j$, the process
\[
 M_n=e^{\theta(ncH-\sum_{m=1}^nS_m)}v_\theta(J_n)
\]
is a nonnegative supermartingale with $M_0=v_\theta(j)$, because
$e^{\theta cH}K_S(\theta)v_\theta\leq v_\theta$.
Its maximal inequality~\cite{williams1991probability} gives, for $y\geq0$,
\[
 \mathbb P(D_S^{(c)}>y\mid J_0=j)
 \leq\min\{1,v_\theta(j)e^{-\theta y}\}.
\]
The continuous-time supremum equals the block-boundary supremum
because service is constant-rate within each coherence block.
This bound is the tail of
$Z_{c,\theta,j}=\theta^{-1}\log v_\theta(j)+E_\theta$,
where $E_\theta\sim\operatorname{Exp}(\theta)$.

Set $x_c=C_G(c)^{1/p}$ and choose $X_c$, independently of
$Z_{c,\theta,j}$, to have the Pareto tail
\[
 \mathbb P(X_c>x)=g_c(x)=
 \begin{cases}
  1, & x\leq x_c,\\
  (x_c/x)^p, & x>x_c.
 \end{cases}
\]
The arrival and conditional service bounds above therefore give
\begin{equation}
\begin{aligned}
 W_A^{(c)}&\leq_{\mathrm{st}}X_c,\\
 \bigl(D_S^{(c)}\mid J_0=j\bigr)
 &\leq_{\mathrm{st}}Z_{c,\theta,j}.
\end{aligned}
\label{eq:app_ppb_stochastic_domination}
\end{equation}
Here $\leq_{\mathrm{st}}$ denotes ordering of the corresponding
tail probabilities at every threshold. Combining
\eqref{eq:app_ppb_pathwise_split} and
\eqref{eq:app_ppb_stochastic_domination} with conditional
independence and closure of the usual stochastic order under independent
addition~\cite{shaked2007stochastic} yields
\[
 \mathbb P(W_0>x\mid J_0=j)
 \leq\mathbb P(X_c+Z_{c,\theta,j}>x).
\]
Fix $c,\theta,j$, so that
$a_{\theta,j}=\theta^{-1}\log v_\theta(j)$ is deterministic.
Write $f_{E_\theta}(u)=\theta e^{-\theta u}$, $u\geq0$, for
the density of $E_\theta$. Conditioning on its value gives
\begin{equation}
\begin{aligned}
 &\mathbb P(X_c+Z_{c,\theta,j}>x)\\
 &\quad=\int_0^\infty
   \mathbb P\bigl(X_c>x-a_{\theta,j}-u\mid E_\theta=u\bigr)
   f_{E_\theta}(u)\,du\\
 &\quad=\int_0^\infty
   \mathbb P(X_c>x-a_{\theta,j}-u)
   f_{E_\theta}(u)\,du\\
 &\quad=\int_0^\infty g_c(x-a_{\theta,j}-u)
                   \theta e^{-\theta u}\,du.
\end{aligned}
\label{eq:app_ppb_convolution_integral}
\end{equation}
The first equality applies the continuous law of total probability
after substituting $E_\theta=u$. The second uses independence,
which leaves the law of $X_c$ unchanged. The last uses the defining
tail $\mathbb P(X_c>y)=g_c(y)$ and the exponential density.
Finally, retain a fixed finite nonempty family of feasible pairs
\[
\begin{aligned}
 \mathcal C\subseteq\{(c,\theta):\;&r<c<\bar s,\quad\theta>0,\\
 &e^{\theta cH}\lambda_S(\theta)\leq1\}.
\end{aligned}
\]
Taking the pointwise minimum of
\eqref{eq:app_ppb_convolution_integral} over these pairs proves
\eqref{eq:ppb_convolution_workload_bound}.
Each member is nonincreasing in $x$ and vanishes at infinity, so
the family minimum is a statewise analytical workload bound.

\bibliographystyle{IEEEtran}
\bibliography{references}

\end{document}